\documentclass[twocolumn]{aastex631}

\usepackage{pifont}
\usepackage{bm}
\usepackage{soul}
\setstcolor{black} 

\newcommand{\Range}[2]{%
  \makebox[3em][c]{#1},\makebox[3em][c]{#2}%
}

\newcommand{\BracketRange}[2]{
  [\Range{#1}{#2}]
}

\newcommand{\StretchedColumn}[3]{
\begin{minipage}[t][#1][t]{#2}
\centering
\vspace*{-0.5\baselineskip}
#3
\vspace*{0.5\baselineskip}
\end{minipage}
}

\soulregister\cite7
\soulregister\ref7
\soulregister\eqref7
\soulregister\ac7
\soulregister\gls7
\soulregister\textcite7
\usepackage{multirow}
\usepackage{longtable,tabu}

\newcommand{\be}{\begin{equation}}
\newcommand{\ee}{\end{equation}}
\newcommand{\bea}{\begin{eqnarray}}
\newcommand{\eea}{\end{eqnarray}}

\usepackage{amsmath}
\usepackage{threeparttable}
\usepackage{xcolor}
\usepackage[normalem]{ulem}

\definecolor{customorange}{HTML}{dd5400}

\definecolor{color1}{HTML}{440154}
\definecolor{color2}{HTML}{481568}
\definecolor{color3}{HTML}{482677}

\definecolor{color18}{HTML}{B8DE29}
\definecolor{color19}{HTML}{DCE318}
\definecolor{color20}{HTML}{FDE725}
\usepackage{tabularx}

\shorttitle{CEDF Models for NS EOS: A Bayesian Cross-Comparison}
\shortauthors{Cartaxo et al.}

\graphicspath{{./}{figures/}}

\begin{document}
\title{Covariant Energy Density Functionals for Neutron Star Matter Equation of State Modeling: Cross-Comparison Analysis Using \texttt{CompactObject}}

\author[0009-0001-7105-8272]{João Cartaxo}
\email{wab.joaocartaxo@gmail.com}
\affiliation{CFisUC, Department of Physics, University of Coimbra, 3004-516 Coimbra, Portugal}

\author[0000-0001-6406-1003]{Chun Huang}
\email{chun.h@wustl.edu}
\affiliation{Physics Department and McDonnell Center for the Space Sciences, Washington University in St. Louis, MO 63130, USA}

\author[0000-0003-2633-5821]{Tuhin Malik}
\affiliation{CFisUC, Department of Physics, University of Coimbra, 3004-516 Coimbra, Portugal}

\author[0000-0002-0169-4003]{Shashwat Sourav}
\affiliation{Physics Department and McDonnell Center for the Space Sciences, Washington University in St. Louis, MO 63130, USA}

\author[0000-0003-2771-759X]{Wen-Li Yuan}
\affiliation{School of Physics and State Key Laboratory of Nuclear Physics and Technology, Peking University, Beijing 100871, China}

\author[0009-0000-8504-9134]{Tianzhe Zhou}
\affiliation{Department of Physics, Tsinghua University, Beijing 100084, China}

\author[0009-0008-3286-7254]{Xuezhi Liu}
\affiliation{Physics Department, Central China Normal University, Luoyu Road, 430030, Wuhan, China}

\author[0000-0001-6464-8023]{Constança Providência}
\affiliation{CFisUC, Department of Physics, University of Coimbra, 3004-516 Coimbra, Portugal}

\begin{abstract}
This study analyzes and contrasts different phenomenological methods used to model the nuclear equation of state (EOS) for neutron star matter based on covariant energy density functionals (CEDF). Using two complementary methodologies, we seek to capture a comprehensive picture of the potential behaviors of ultra-dense nucleonic matter and identify the most plausible models based on current observational and experimental constraints. Observational data from radio pulsar timing, gravitational wave detection of GW170817, and X-ray timing provide critical benchmarks for testing the models. We have derived the EOS posteriors for various CEDF models within the \texttt{CompactObject} package, utilizing recent observational data on neutron stars, state-of-the-art theoretical constraints from chiral effective field theory ($\chi$EFT) calculations for pure neutron matter at low densities, and pQCD-derived constraints. Our analysis has demonstrated that while all considered CEDF models broadly reproduce current astrophysical and theoretical constraints, subtle yet important differences persist among them, with each framework exhibiting distinct characteristics at supra-nuclear density. This is in particular true for the proton fraction inside neutron stars, but also supported by the models' behavior with respect to the pure neutron matter EOS and the density dependence of the speed of sound. Our study highlights the sensitivity of dense matter predictions to the underlying EOS parameterizations and the priors considered.
\end{abstract}

\section{Introduction}  
\label{sec:intro}
Neutron stars, which are remnants of massive stellar explosions, are among the most extreme objects in the universe \citep{Fryer:2013vja}. These compact astrophysical laboratories provide a unique opportunity to study matter under conditions that far exceed those achievable in terrestrial experiments \citep{book.Glendenning,Burgio:2021vgk}. With densities up to {7-8} times greater than those found in atomic nuclei, neutron stars offer an unparalleled environment for testing our understanding of the equation of state (EOS) of ultra-dense matter \citep{book.Haensel2007,Tan:2020ics,Tan:2021nat, Marczenko:2022cwq, Altiparmak:2022bke}. The EOS, which describes how pressure relates to density, is crucial for determining the structure and evolution of neutron stars, as well as their observable properties, including mass, radius, and {tidal deformability} \citep{Imam:2021dbe,Malik:2022ilb,Coughlin:2019kqf,Wesolowski:2015fqa,Furnstahl:2015rha,Ashton2019,Landry:2020vaw,CompactObject,Char:2023fue,Imam:2024gfh,2024MNRAS.529.4650H,2025MNRAS.536.3262H}.

The extreme conditions within neutron star cores may give rise to a rich spectrum of exotic degrees of freedom, ranging from hyperonic matter \citep{Ambartsumyan60,Glendenning:1982nc,Glendenning:1984jr,Glendenning:1991es,Schaffner:1995th} to deconfined quarks \citep{Haensel:1986qb,Alcock:1986hz} and even potential dark-matter admixtures \citep{PhysRevD.77.043515} may emerge in their core region. Additionally, insights into the EOS significantly contribute to nuclear physics by bridging the gap between laboratory experiments at subnuclear densities and extreme conditions in neutron stars \citep{1998PhRvC..58.1804A,2024NatAs...8..328T}. This interplay between astrophysics and nuclear physics highlights the importance of studying the EOS as a unifying framework for understanding matter across vastly different scales.

In this comprehensive analysis, we aim to dissect and compare various methods for modeling the EOS of neutron star matter within the phenomenological approach. EOS modeling approaches generally fall into two broad categories: phenomenological models based on covariant energy density functionals (CEDF) and model-agnostic EOS descriptions. Phenomenological {CEDF} models are grounded in covariant theoretical frameworks \cite{Boguta1977,Sugahara:1993wz,Mueller1996,Typel1999,Lalazissis2005,Todd-Rutel2005} (see \cite{Dutra:2014qga} for a review). These models provide detailed predictions about the behavior of matter under extreme conditions but rely heavily on assumptions tied to their underlying theories. On the other hand,  model-agnostic EOS adopt a more flexible approach, designed to explore the parameter space without strong theoretical preconceptions. As model-agnostic EOS, different parametric  (\cite{Hebeler2013,Kurkela:2014vha,Tews:2018kmu,Annala:2019puf,Capano:2019eae} and non-parametric  \cite{Landry:2020vaw,Essick2019,Gorda:2022jvk} descriptions have been considered. A different category of EOS description that will not be considered are nuclear meta-models based on a Taylor expansion of the energy which is parameterized by the empirical nuclear matter parameters \citep{Margueron2018a,Zhang2018}. By combining complementary methodologies, we seek to capture the full range of possible behaviors for ultra-dense matter and identify the most plausible models based on current observational and experimental constraints. As the first part of our series of survey studies, this work focuses exclusively on phenomenological CEDF models.

It is worthwhile to discuss the validity of using a hadronic model to describe neutron star matter. At a certain density, the hadrons will dissolve, and deconfined matter should be considered. In particular, this should occur when the hard cores of the nucleons (hadrons) begin to overlap.  According to \cite{Kaiser:2024vbc}, a spectral analysis of the isoscalar electric charge form factor, axial form factor, and mass form factor, the mean-square radius of the nucleon core (which contains the three quarks that define the baryon number) is approximately half a fermi. They obtained the following values: $0.50\pm0.01$ fm, $\approx 0.53$ fm, and $0.48\pm0.05$ fm for the hard core obtained from the isoscalar electric form factor, the form factor associated with the isovector axial current, and the mass radius, respectively. As discussed in \cite{Kaiser:2024vbc},  this means that the quark cores will start to overlap for a baryonic density $\rho\gtrsim 6\rho_0$, the close packing of hard spheres with a 0.5~fm radius giving $\rho\sim 8 \rho_0$. Therefore, we consider the density range in the present study to be acceptable if deconfinement is defined by nucleon core overlap.

To rigorously evaluate these models, we employ Bayesian evidence as a model selection tool \citep{kass}. This statistical framework allows us to weigh the predictive power of each {EOS model} against astrophysical observations and nuclear physics experiments. Observational data from {radio} pulsar timing {(eg. \cite{Fonseca_2021,Reardon_2024})}, gravitational wave {detection of double neutron star merger event GW170817 \citep{LIGOScientific:2017vwq}}, and X-ray observations of neutron stars {(e.g. \cite{Riley:2019yda,Miller2019,Choudhury:2024xbk,Dittmann24})} provide critical benchmarks for testing the models. In addition, constraints from nuclear physics experiments at lower densities ensure that our models remain consistent with well-understood properties of matter under less extreme conditions.{(e.g. \cite{crex,prex})}

The analysis presented in this work is facilitated by a fully open-source Bayesian inference software package developed by the authors \citep{CompactObject}. This tool is specifically designed to constrain the neutron star EOS by integrating inputs from both astrophysical observations and nuclear experiments. By making our methodology publicly available, we aim to foster collaboration within the broader scientific community and to encourage further advancements in this field.

In this work, we provide a comprehensive Bayesian analysis that systematically compares multiple categories of  {CEDF} models using identical observational constraints and computational frameworks, enabling direct comparison of their Bayes factors. Our analysis incorporates the latest observational data, including recent NICER measurements of PSR J0437+4715, while integrating theoretical limits from $\chi$EFT (at low densities) and pQCD (at high densities) with multi-messenger astronomical constraints. As a significant contribution to research infrastructure, we release our complete posterior distributions -- approximately {145,000} samples across all models -- as an open-source dataset with a dedicated graphical interface tool, enabling community applications in machine learning, meta-analyses, and model validation \citep{cartaxo_2025_15482081}. This article constitutes the first part of our broader survey of neutron star EOS, with a future companion paper extending the analysis to {model-agnostic descriptions} that explore parameter space without strong theoretical preconceptions. By utilizing the open-source \texttt{CompactObject} package, we ensure full reproducibility and establish a foundation for future studies incorporating additional degrees of freedom such as hyperons and quarks. 
This comprehensive approach allows us to understand how the different covariant density functionals describe current observational data. It also highlights some specific properties, in particular proton fraction profiles and sound-speed behavior, that can differentiate between otherwise similar models. We expect that such a comparison will improve physically motivated models that describe the EOS of neutron star matter and provide important insights for future observations that could identify specific properties that should be described by the underlying nuclear matter models.  The generalization of the models considered in the present study could include, among others,  adding extra non-linear terms \citep{Mueller1996,Agrawal:2010wg,Li:2022okx}, including the scalar isovector meson $\delta$ \citep{Liu:2001iz,Gaitanos:2003zg}, introducing $\omega$ and $\varrho$-meson tensor couplings \citep{Rufa:1988zz,Typel:2020ozc}, or considering a generalization of the density dependence of the coupling parameters in density dependent models.

The paper is organized as follows: Section \ref{sec:model} outlines the EOS formalism for the various models examined; Section \ref{sec:like} describes our inference framework and the observational datasets employed; Section \ref{sec:results} discusses the resulting posterior distributions and Bayes evidence of each models; and Section \ref{sec:conclusions} offers concluding remarks and suggests directions for future work.
\section{Formalism} 
\label{sec:model}
In this section, we provide a summary of the frameworks of different phenomenological models within the mean field approximation, used to describe the EOS under the \texttt{CompactObject} framework, see \citep{CompactObject}\footnote{\url{https://github.com/ChunHuangPhy/CompactObject}}. It should be noted that the current version of \texttt{CompactObject} supports only nucleonic degrees of freedom for all CEDF models. {We introduce models through their Lagrangian density, employing two distinct approaches for incorporating nuclear interactions. The first approach employs constant nucleon-meson couplings supplemented by non-linear mesonic terms, exemplified by the NL models \citep{Boguta1977,Sugahara:1993wz,Mueller1996,Todd-Rutel2005}.
The second approach utilizes nucleon-meson couplings that vary with density, represented by several models: DDB \citep{2022ApJ...930...17M}, DDH \citep{Typel1999}, and GDFMX \citep{Char:2023fue}. The parametrization GDFMX has been derived  from the  GDFM model \citep{Gogelein:2007qa}. The coupling parametrizations within this model include an $x^3$ term (see Eq. (\ref{gdfm})) that could potentially lead to an {uncontrolled} behavior in the EOS. 
{Note, however, that the original GDFM model includes also the scalar isovector $\delta$-meson, and, in addition, the coupling to the $\rho$-meson has the coefficient of the cubic term set to zero. The parametrization of the couplings in this model was defined to reproduce the Dirac Brueckner Hartree-Fock results obtained in \citep{vanDalen:2006pr}, all couplings showing a decrease with density.} {To obtain a similar behavior, in particular, decreasing couplings with density,} we have implemented a constraint in our Bayesian inference calculations that ensures GDFMX couplings never increase with density.}

\subsection{The Models\label{sec:models}}
The EOS of nuclear matter is determined from the Lagrangian density that describes the nuclear system. The degrees of freedom include the nucleons of mass $m$ described by Dirac spinors ${\Psi}$, and the meson fields, the scalar isoscalar $\sigma$ field, the vector isoscalar $\omega$ field, and the vector isovector $\varrho$ field, with masses $m_i,\, i=\sigma, \, \omega,\, \varrho$, which describe the nuclear interaction. The parameters $\Gamma_i$ or $g_i$ {(for constant couplings)}, $i=\sigma, \, \omega,\, \varrho$ designate the couplings of the mesons to the nucleons.

The Lagrangian density is given by
\begin{equation}
\resizebox{\columnwidth}{!}{$
\begin{aligned}
\mathcal{L} =\; & \bar{\Psi}\Big[\gamma^{\mu}\Big(i\partial_{\mu}-\Gamma_{\omega} 
\omega_{\mu}
-\Gamma_{\varrho}\,\boldsymbol{t}\cdot\boldsymbol{\varrho}_{\mu}\Big)\\[1mm]
& \quad -\Big(m-\Gamma_{\sigma}\sigma\Big)\Big]\Psi 
+ \frac{1}{2}\Big(\partial_{\mu}\sigma\,\partial^{\mu}\sigma-m_{\sigma}^{2}\sigma^{2}\Big)\\[1mm]
& \quad -\frac{1}{4} F_{\mu\nu}^{(\omega)}F^{(\omega)\mu\nu} 
+\frac{1}{2}m_{\omega}^{2}\omega_{\mu} \omega^{\mu}\\[1mm]
& \quad -\frac{1}{4} \boldsymbol{F}_{\mu\nu}^{(\varrho)}\cdot\boldsymbol{F}^{(\varrho)\mu\nu} 
+ \frac{1}{2}m_{\varrho}^{2}\,\boldsymbol{\varrho}_{\mu}\cdot\boldsymbol{\varrho}^{\mu}\\[1mm]
& \quad +\mathcal{L}_{NL}.
\end{aligned}
$}
\label{eq:Lag}
\end{equation}

where $\gamma^\mu$ represents the Dirac matrices, $\boldsymbol{t}$ the isospin operator, and vector meson field strength tensors are defined as $F^{(\omega)\mu \nu} = \partial^\mu \omega^{\nu} -\partial^\nu \omega^{\mu}$ and $F^{(\varrho)\mu \nu} = \partial^\mu \boldsymbol{\varrho}^{ \nu} -\partial^\nu \boldsymbol{\varrho}^{\mu} - \Gamma_\varrho (\boldsymbol{\varrho}^{\mu} \times \boldsymbol{\varrho}^{\nu})$. For models with constant couplings the density dependent couplings $\Gamma_i$ in eq. (\ref{eq:Lag}) should be replaced by constant couplings $g_i$. The term $\mathcal{L}_{NL}$ is absent in models with density dependent couplings, and includes self-interacting and mixed meson terms in models with constant couplings $g_i$. In the following, we identify $\varrho$ with the mean-field time component of the ${\varrho_0}^\mu$ component of the isospin vector field  $\boldsymbol{\varrho}^\mu$, and $\omega$ with the mean-field time component of the vector field $\omega^\mu$.

\subsubsection{Density Dependent Description}
The density dependent models include meson-nucleon couplings $\Gamma_i$ that depend on the total nucleonic density $\rho$:
\begin{equation}
  \Gamma_{i}(\rho) =\Gamma_{i,0} ~ h_i(x)~,\quad x = \rho/n_0~, \,i=\sigma, \omega, \varrho, 
\end{equation}
with $\Gamma_{i,0}$ the couplings at saturation density $n_0$. For the isoscalar mesons, $\sigma$ and $\omega$, two parametrizations $h_i$ are considered:
\begin{equation}
h_i(x) = \exp[-(x^{a_i}-1)]
\end{equation}
as in DDB sets, and
\begin{eqnarray}
h_i(x) =a_M\frac{1+b_i(x+d_i)^2}{1+c_i(x+d_i)^2} \, ,
\end{eqnarray}
for DDH data sets.

The $\varrho$-meson nucleon coupling is defined as:
\begin{equation}
h_\varrho(x) = \exp[-a_\varrho (x-1)] ~.
\label{grho}
\end{equation}
We will also discuss the effect of including an extra parameter in the above definition
\begin{equation}
    h_\varrho(x) = y\, \exp[-a_\varrho (x-1)] + (y-1), \quad 0 < y \leq 1.
\label{y}
\end{equation}
Models with this extra parameter are designated by 'model'$y$.
For the GDFMX set, the density dependence of the  couplings follows \citep{Char:2023fue}:
\begin{equation}
\Gamma_i(\rho) = a_i + \bigl(b_i + d_i \, x ^3\bigr)\, e^{- c_i x},
\label{gdfm}
\end{equation}

where \(i = \sigma, \omega, \varrho\) and \(x = \frac{\rho}{n_0}\). Here \(n_0\) is a normalization density ({$n_0 = 0.16$ fm$^{-3}$}). This approach is based on the one proposed in \citep{Gogelein:2007qa} but removing the correction term for the $\omega$-meson at saturation.

Note that for the DDH coupling scheme, see Ref. \citep{Typel1999}, additional constraints are introduced that reduce the number of free parameters; these are as follows: $h_i(1) = 1$, $h_i^{\prime\prime}(0) = 0$, and $h_\sigma^{\prime\prime}(1) = h_\omega^{\prime\prime}(1)$.

\subsubsection{Non-linear Meson Terms}
The NL models employ constant couplings $g_i$ and include non-linear meson terms in the Lagrangian density:
\begin{equation}
\resizebox{\columnwidth}{!}{$
\begin{aligned}
\mathcal{L}_{NL} =\; & -\frac{\kappa}{3!} (g_\sigma\sigma)^3 
-\frac{\lambda}{4!} (g_\sigma\sigma)^4 \\[1mm]
&+\frac{\xi}{4!}\Bigl(g_{\omega}^2\,\omega_{\mu}\omega^{\mu}\Bigr)^2 
+\Lambda_{\omega}\, g_{\varrho}^{2}\,\boldsymbol{\varrho}_{\mu}\cdot\boldsymbol{\varrho}^{\mu}\, g_{\omega}^2\,\omega_{\mu}\omega^{\mu}\,.
\end{aligned}
$}
\end{equation}

The parameters $\kappa$, $\lambda$, $\xi$, and $\Lambda_\omega$ control various aspects of nuclear matter:
\begin{itemize}
\item The parameters $\kappa$ and $\lambda$ affect  the nuclear matter incompressibility at saturation \citep{Boguta1977};
\item $\xi$ modulates the high density EOS behavior (larger $\xi$ produces softer EOS at high densities) \citep{Sugahara:1993wz};
\item The $\omega-\varrho$ term influences the density dependence of the symmetry energy \citep{Todd-Rutel2005}.
\end{itemize}

\subsubsection{Equations of motion}

For RMF-NL models, the equations of motion for the meson fields are given by:
\begin{eqnarray}
	{\sigma}&=& \frac{g_{\sigma}}{m_{\sigma,{\rm eff}}^{2}}\sum_{i} \rho^s_i\\
	{\omega} &=&\frac{g_{\omega}}{m_{\omega,{\rm eff}}^{2}} \sum_{i} \rho_i \\
	{\varrho} &=&\frac{g_{\varrho}}{m_{\varrho,{\rm eff}}^{2}}\sum_{i} t_{3i} \rho_i,
\end{eqnarray}
where $\rho^s_i$ and $\rho_i$ are, respectively, the scalar density and the number density of nucleon $i$ and  the effective meson masses are defined as:
\begin{eqnarray}
   m_{\sigma,{\rm eff}}^{2}&=& m_{\sigma}^{2}+{ \frac{\kappa}{2} g_\sigma^3}{\sigma}+{\frac{\lambda}{6} g_\sigma^4}{\sigma}^{2} \\ 
   m_{\omega,{\rm eff}}^{2}&=& m_{\omega}^{2}+ \frac{\xi}{3!}g_{\omega}^{4}{\omega}^{2} +2\Lambda_{\omega}g_{\varrho}^{2}g_{\omega}^{2}{\varrho}^{2}\\
   m_{\varrho,{\rm eff}}^{2}&=&m_{\varrho}^{2}+2\Lambda_{\omega}g_{\omega}^{2}g_{\varrho}^{2}{\omega}^{2}.
\end{eqnarray}

The non-linear meson terms create density-dependent effects:
\begin{itemize}
\item $m_{\omega,{\rm eff}}$ increases with the $\omega$-field, making the $\omega$ field increase  with density with a power of $\rho$ {below one. If $\xi =0$, $\omega\propto  \rho$}. 
\item $m_{\varrho,{\rm eff}}$ increases with density, weakening the $\varrho$ field at higher densities and softening symmetry energy.
\end{itemize}

For density dependent models (DDH, DDB and GDFM), these equations still apply with $g_i=\Gamma_i$ and  $m_{i,{\rm eff}} = m_i$, as non-linear terms are absent, i.e.
\begin{eqnarray}
	{\sigma}&=& \frac{\Gamma_{\sigma}}{m_{\sigma}^{2}}\sum_{i} \rho^s_i\\
	{\omega} &=&\frac{\Gamma_{\omega}}{m_{\omega}^{2}} \sum_{i} \rho_i \\
	{\varrho} &=&\frac{\Gamma_{\varrho}}{m_{\varrho}^{2}}\sum_{i} t_{3i} \rho_i.
\end{eqnarray}

{\subsection{Crust Model definition}
Following the prescription of \cite{2024MNRAS.529.4650H,2025MNRAS.536.3262H}, we construct a unified EOS by appending the BPS outer-crust EOS to a $4/3$-polytropic inner-crust segment and matching this smoothly to the core EOS at the crust–core transition density
$\varepsilon_{c}\simeq2.14\times10^{14}\,\mathrm{g\,cm^{-3}}$.  The
polytropic parameters $A$ and $B$ are fixed by demanding continuity of
both $P$ and $P'\!(\varepsilon)$ at
$\varepsilon_{\text{outer}}=4.3\times10^{11}\,\mathrm{g\,cm^{-3}}$ and
$\varepsilon_{c}$.  The
resulting piecewise form  
\[
P(\varepsilon)=
\begin{cases}
P_{\mathrm{BPS}}(\varepsilon), & \varepsilon_{\min}\le\varepsilon\le\varepsilon_{\text{outer}},\\
A + B\,\varepsilon^{4/3},      & \varepsilon_{\text{outer}} < \varepsilon \le \varepsilon_{c},\\
P_{\mathrm{RMF}}(\varepsilon), & \varepsilon_{c} < \varepsilon,
\end{cases}
\]
ensures a monotonically increasing pressure over the entire density
range required to solve the TOV equations.  Full derivation and
numerical details are given in \cite{2024MNRAS.529.4650H,2025MNRAS.536.3262H}.}

\section{
Inference Framework}
\label{sec:like}
{This section provides a detailed description of the inference framework implemented in our study. Specifically, it focuses on defining the likelihood we are using in our Bayesian analyses. For the prior definition, we keep consistence with several previous studies based on similar models (e.g. \cite{2022ApJ...930...17M,Beznogov:2022rri,2024MNRAS.529.4650H,Malik:2023mnx}), for more details see Table \ref{tab:prior}.} In this study, we aim to constrain the nuclear EOS parameters, denoted by $\boldsymbol{\theta}$, by combining independent constraints from nuclear experiments, gravitational-wave observations, and X-ray timing data. Assuming that the various data sets are statistically independent, the overall log-likelihood is the sum of the individual log-likelihoods associated with each data set. In what follows we describe each component.

\textbf{Nuclear Matter Properties (NMP):}  
For the nuclear saturation properties, we adopt a Gaussian likelihood, which reads
\begin{equation}
    \mathcal{L}^{\rm NMP}(\boldsymbol{\theta}) = \prod_{j=1}^{N_{\rm NMP}}  \frac{1}{\sqrt{2\pi\, \sigma_j^2}} e^{- \frac{1}{2}\left(\frac{D_j - m_j(\boldsymbol{\theta})}{\sigma_j}\right)^2},
    \label{eq:gausslikelihood}
\end{equation}
where the index $j$ runs over the $N_{\rm NMP}$ independent observables (with $D_j$ denoting the measured value, $m_j(\boldsymbol{\theta})$ the model prediction, and $\sigma_j$ the uncertainty). The nuclear matter properties considered include the saturation density $n_0 = 0.153 \pm 0.005$ fm$^{-3}$, the binding energy per nucleon $\epsilon_0 = -16.0 \pm 0.2$ MeV, the incompressibility $K_0 = 230 \pm 40$ MeV \citep{Margueron:2012mw}, and the symmetry energy $J_{\rm sym,0} = 32.5 \pm 2.5$ MeV. Note that at $2\sigma$ the values considered for $K_0$ cover the wider range of values discussed in the literature 175-255 MeV\citep{Huth:2020ozf} and 250-315 MeV \citep{Stone:2014wza}.

\textbf{Astrophysical Constraints:}  
We incorporate two types of astrophysical observations, each with its own likelihood function.

\emph{Tidal Deformability from GW170817:}  
Following the LIGO-Virgo Collaboration \citep{LIGOScientific:2018cki}, we use gravitational-wave data from the binary neutron star merger GW170817 \citep{Abbott_2017,Abbott_2018}. The likelihood for the tidal deformability is defined as
\begin{equation}
    \mathcal{L}^{\rm GW170817}(\boldsymbol{\theta}) = \prod_{i} p\Bigl( \Lambda_{1,i}, \Lambda_{2,i}, q_{i} \,\bigl|\, \mathcal{M}_{c}, \boldsymbol{d}_{\text{GW},i}\, (\boldsymbol{d}_{\text{EM},i}) \Bigr),
    \label{eq:gwlikelihood}
\end{equation}

where the product is over posterior samples indexed by $i$. Here, $\Lambda_{1,i}$ and $\Lambda_{2,i}$ denote the tidal deformabilities of the binary components, $q_i$ is the mass ratio, $\mathcal{M}_c$ is the chirp mass, and $\boldsymbol{d}_{\text{GW},i}$ (optionally supplemented by electromagnetic data $\boldsymbol{d}_{\text{EM},i}$) represents the observational data. {A similar pipeline for modeling the GW-related likelihood is presented in detail by \citep{Raaijmakers_2021} and \citep{2024MNRAS.529.4650H}, and is therefore omitted here.}

\emph{Masses and Radii from NICER:}  
NICER’s X-ray {spectral-timing} observations provide independent constraints on neutron star masses and radii. The likelihood associated with these measurements is given by
\begin{equation}
    \mathcal{L}^{\rm NICER}(\boldsymbol{\theta}) = \prod_{j} p\Bigl( M_{j}, R_{j} \,\bigl|\, \boldsymbol{d}_{\text{NICER},j} \Bigr),
    \label{eq:nicerlikelihood}
\end{equation}
where the product runs over the NICER data sets (indexed by $j$), with $M_j$ and $R_j$ being the measured mass and radius, and $\boldsymbol{d}_{\text{NICER},j}$ the corresponding NICER observational data. 
{This study implemented three Mass-Radius (MR) posterior informed from NICER observations. The first one is NICER observations of PSR J0437–4715 (J0437) \citep{Choudhury:2024xbk}, whose mass is precisely determined to be $1.44\pm0.07\,M_\odot$ via radio timing \citep{ Reardon_2024}. The second source is PSR J0030+0451 (J0030), initial NICER data lacked a direct mass constraint until \cite{Riley:2019yda} and \cite{Miller2019} reported masses of $1.34_{-0.16}^{+0.15}\,M_\odot$ and $1.44_{-0.14}^{+0.15}\,M_\odot$, respectively, alongside radius estimates. Subsequent analyses demonstrate that J0030’s inferred mass–radius relation is highly sensitive to hotspot geometry \cite{Vinciguerra_2024}. In this study, we adopt the MR posterior distribution from the ST+PDT hotspot model presented by \citep{Vinciguerra_2024}, which aligns well with gamma-ray–inferred magnetic field geometries \citep{2021ApJ...907...63K}. To mitigate these systematic uncertainties of the geometry of the hotspot, the implementation of more physics-motivated hotspot configurations could be one possible solution \citep{Huang:2025jyk}. We also implemented the  mass-radius measurement of PSR J0740+6620 (J0740) as a constraint. We utilized the most up-to-date MR measurement of radius $12.49_{-0.88}^{+1.28}$ km and mass $2.073_{-0.069}^{+0.069} M_{\odot}$ star reported by \citep{salmi2024}.
}

\textbf{Theoretical Constraints:}  
Additional constraints are imposed to ensure consistency with theoretical expectations at both high and low densities. These include:

\emph{pQCD-derived Constraints:}  
At high densities, perturbative QCD (pQCD) calculations provide predictions for certain thermodynamic quantities.
{Although the densities involved in the pQCD calculations are very high, $\sim 40 n_0$, \cite{komoltsev_2023_7781233} have shown that causality and the  thermodynamical relations 
$$ c^{-2}_s=\frac{\mu_B}{n_B} \frac{\partial n_B}{\partial \mu_B}\ge 1, \quad  \int_{\mu_L}^{\mu_H} n_B(\mu_B) d\mu_B=p_H-p_L,$$
where $p_L$ and $p_H$ are the pressure at low and high densities and $B$ refers to baryonic, would impose constraints on the densities attained inside a neutron star. In \cite{komoltsev_2023_7781233}, the authors have considered for the low limit the results from $\chi$EFT \cite{Hebeler2013} and for the high limit the pressure obtained within a pQCD calculation.
}
We impose these constraints via
\begin{equation}
    \mathcal{L}(d_{pQCD} |\boldsymbol{\theta}) = P(d_{pQCD}|\boldsymbol{\theta}) = \mathcal{L}^{\rm pQCD}(\boldsymbol{\theta}).
    \label{eq:pqcdlikelihood}
\end{equation}
{where $d_{pQCD}$ represents the probability distribution in the energy density–
pressure plane at $\rho = 1.2$ fm$^{-3}$ (the highest density point of the calculated EOS), utilizing the likelihood function from
\cite{komoltsev_2023_7781233}.}

{\textbf{\it Choice of Prior on Renormalization Scale:} The original pQCD implementation
by \cite{komoltsev_2023_7781233} employed a uniform prior on the renormalization
scale $X$ in logarithmic space (i.e., uniform on $\log X$), which when transformed
to linear space yields a probability distribution that assigns greater weight to
lower values of $X$ and diminished weight to higher values. Specifically, a
uniform prior on $\log X$ over the interval $[\log(1), \log(4)]$ corresponds to
a probability density $p(X) \propto 1/X$ in linear space, thus favoring smaller
renormalization scales.}

{We find no compelling physical justification for this choice. In the absence of prior knowledge favoring specific scales within the theoretically motivated range, the principle of maximum entropy and the standard convention in renormalization group analysis suggest adopting a uniform prior on the renormalization scale $X$ in linear space over the interval $[1,4]$. This ensures equal a priori probability for all scales within the physically reasonable range.} 

{Consequently, for all main results presented in this study, we employ a uniform prior on $X$ (in linear space) when sampling the pQCD constraint at $\rho = 1.2$ fm$^{-3}$, which corresponds to the maximum density point of the calculated EOS. The sensitivity of our results to this prior choice is examined in Appendix, where we compare posteriors obtained using the uniform-linear prior versus the uniform-logarithmic prior (Figure \ref{fig:pQCD prior1}).}

\emph{$\chi$EFT-derived Constraints:}
Below saturation density, chiral effective field theory ($\chi$EFT) provides robust constraints for pure neutron matter. In our \texttt{CompactObject} package, the likelihood function for $\chi$EFT data allows users to choose either energy or pressure constraints using the ``e" or ``p" tag, respectively. It also offers the option to select between a Gaussian or Super-Gaussian likelihood, with an additional feature to enlarge the uncertainty band see documentation \citep{CompactObject}. 
{In this study, we adopt a Super-Gaussian likelihood and the available band obtained from various $\chi$EFT calculations compiled in Ref. \citep{Huth:2021bsp}.}  

The likelihood is given by
\begin{equation}
    \mathcal{L}^{\rm PNM}(D \mid \boldsymbol{\theta}) = \prod_j \frac{1}{2 \sigma_j^2} \frac{1}{\exp\left(\frac{|D_j - m_j(\boldsymbol{\theta})| - \sigma_j}{\Delta}\right) + 1}.
    \label{eq:boxlikelihood}
\end{equation}
where $D_j$ is the median value and $\sigma_j$ corresponds to the uncertainty of the $j^{\rm th}$ data point from the energy per neutron $\chi$EFT constraints. The value of $\Delta$ is $0.015$, which is quite reasonable for our purposes. 
{We consider these constraints at $\rho=0.04$~fm$^{-3}$, $\rho=0.08$~fm$^{-3}$,  $\rho=0.12$~fm$^{-3}$ and   $\rho=0.16$~fm$^{-3}$.}

\textbf{Total Likelihood:}  
Assuming that the different data sets are independent, the overall likelihood is given as:
\begin{equation}
    \begin{split}
        \mathcal{L}_{\rm total}(\boldsymbol{\theta}) & = \mathcal{L}^{\rm NMP}(\boldsymbol{\theta}) \times \mathcal{L}^{\rm GW170817}(\boldsymbol{\theta}) \\
        & \times \mathcal{L}^{\rm NICER}(\boldsymbol{\theta}) \times \mathcal{L}^{\rm pQCD}(\boldsymbol{\theta}) \times \mathcal{L}^{\rm \chi EFT}(\boldsymbol{\theta}).
    \end{split}
    \label{eq:totallikelihood}
\end{equation}
This combined likelihood serves as the foundation for our Bayesian inference of the EOS parameters $\boldsymbol{\theta}$, rigorously incorporating uncertainties from both observational and theoretical inputs.

\vfill

\section{Results} \label{sec:results}
In this section we present our main results. We discuss the EOS posteriors in Sec. \ref{subsec:EOS}, the mass-radius and $\Lambda$-mass probability distributions in Sec. \ref{subsec:mr}, the proton fraction in Sec.\ref{subsec:yp}, the speed of sound distributions in Sec. \ref{subsec:cs} and finally we compare the Bayes factors of the different models introduced in Sec. \ref{subsec:bayesf}.

\begin{figure*}
    \centering
    \includegraphics[width=1.0\linewidth]{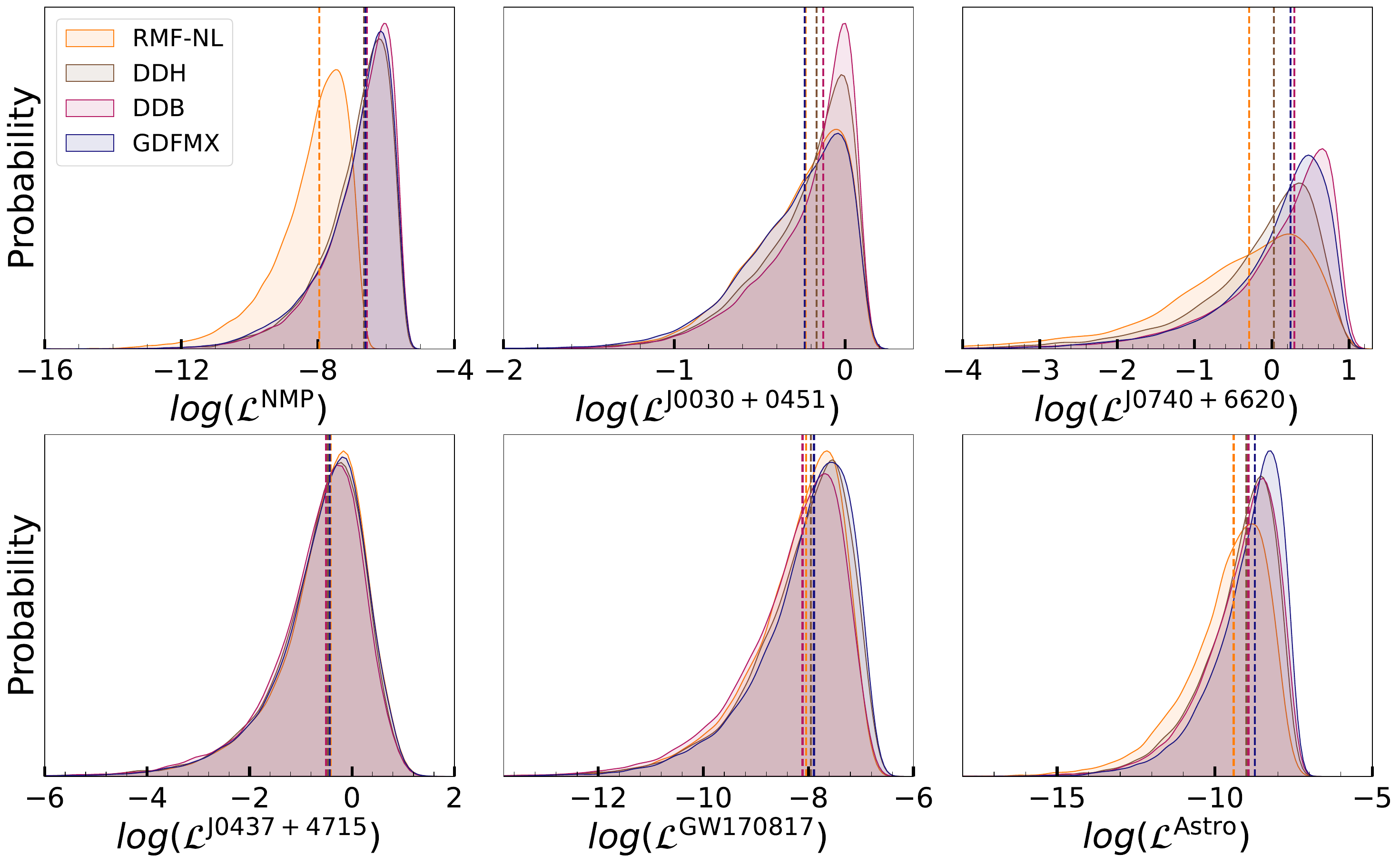}
    \caption{Log-likelihood distributions of the posterior obtained for RMF-NL, DDH, DDB, and GDFMX models across six panels: (i) Contributions from nuclear saturation properties, $log(\mathcal{L}^{\rm NMP})$; (ii) MR constraints from NICER measurements of PSR J0030+0451 \citep{Riley:2019yda}, $log(\mathcal{L}^{\rm J0030+0451})$; (iii) MR constraints for PSR J0740+6620 \citep{Riley:2021pdl}, $log(\mathcal{L}^{\rm J0740+6620})$; (iv) MR constraints for PSR J0437+4715 \citep{Choudhury:2024xbk}, $log(\mathcal{L}^{\rm J0437+4715})$; (v) Dimensionless tidal deformability constraints from GW170817 \citep{LIGOScientific:2018cki}, $log(\mathcal{L}^{\rm GW170817})$; (vi) Total astrophysical contribution combining all NICER and GW170817 data, $log(\mathcal{L}^{\rm Astro})$. The vertical dashed lines, each in its respective color, signify the median of the distributions for each model.}
    \label{fig:loglike}
\end{figure*}
\subsection{EOS posterior\label{subsec:EOS}}
{We have derived the EOS posteriors for various phenomenological EOS models within
the \texttt{CompactObject} framework, using recent observational data on neutron
star properties, state-of-the-art theoretical constraints from $\chi$EFT
calculations for pure neutron matter at low densities ($\rho \leq 0.16$ fm$^{-3}$),
and pQCD-derived constraints applied at the high-density anchor point
$\rho = 1.2$ fm$^{-3}$ (approximately $7.5 n_0$), which corresponds to the maximum density point of the calculated EOS.} These constraints, along with some minimal nuclear matter saturation properties (NMP), were discussed above in Sec. \ref{sec:like}. In this section, we will examine and contrast the posterior distributions of various EOS properties obtained for the EOS models under consideration, specifically NL, DDB, DDH, and GDFMX.

\begin{figure}[ht!]
    \centering
    \includegraphics[width=1.0\linewidth]{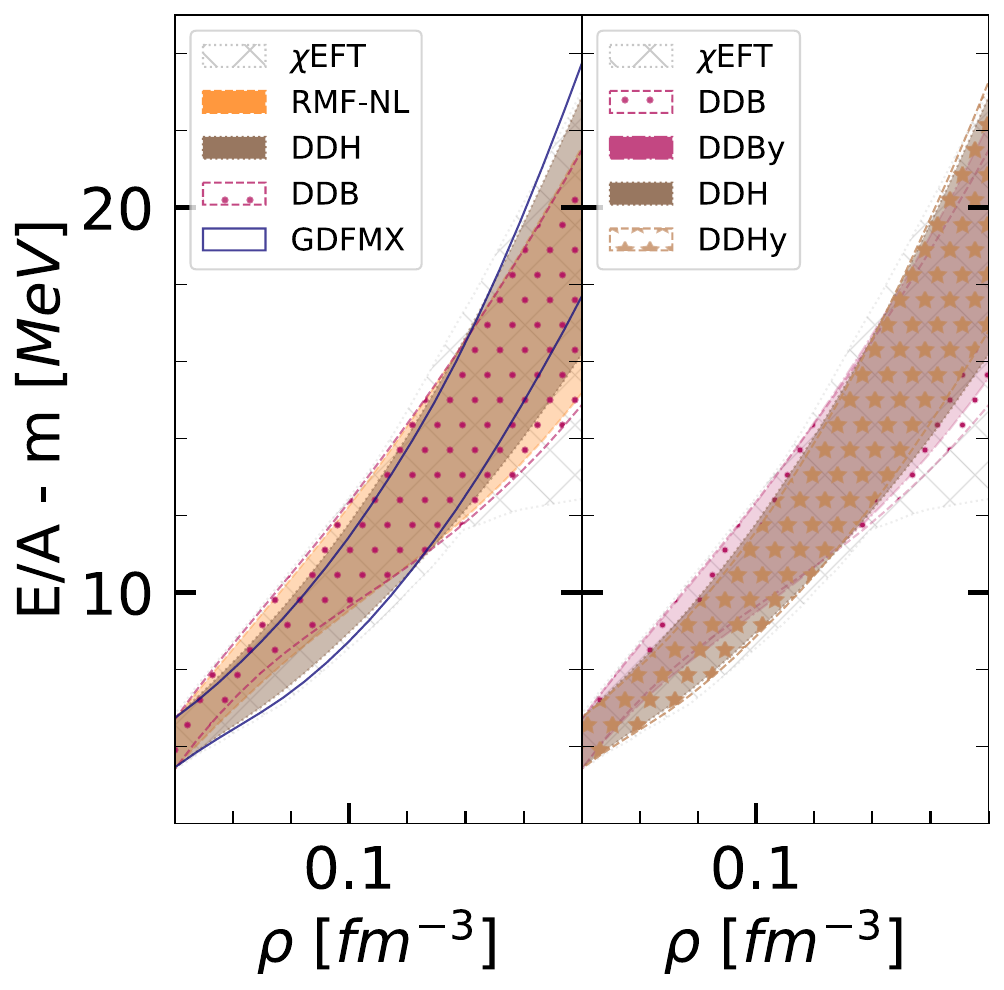}
    \caption{The 90\% credible interval (CI) of binding energy (E/A - m, where m is nucleon mass) for pure neutron matter versus number density $\rho$ for phenomenological CEDF models in this study, compared with $\chi$EFT constraints from multiple sources (Ref. \citep{Huth:2021bsp}). Density dependent models labeled 'model'$y$ include an extra parameter in the definition of the coupling $\Gamma_\varrho$ as defined in Eq. (\ref{y}).}
    \label{fig:EOS_pnm}
\end{figure}
In Figure~\ref{fig:loglike}, we show the posterior log-likelihood distributions for four phenomenological EOS models -- {RMF-NL (light orange), DDH (tan), DDB (pink), and GDFMX (blue)} -- across six panels corresponding to distinct constraints with mostly Gaussian likelihoods \footnote{The constraints of $\chi$-EFT have been integrated using super-Gaussian probabilities that account for the binding energy (E/A - m, with m being the nucleon mass) of pure neutron matter, drawing from various sources gathered in Ref \citep{Huth:2021bsp}. In addition, the pQCD implementation uses a publicly accessible likelihood function from \citep{komoltsev_2023_7781233}.}: nuclear saturation properties, mass-radius data for PSR J0030+0451, PSR J0740+6620, and PSR J0437+4715, tidal deformability of GW170817, and the combined astrophysical dataset. The x-axis depicts $\log(\mathcal{L})$, while the y-axis shows the associated probability density, with vertical dashed lines indicating each model’s median log-likelihood. 

Overall, all models perform comparably well and show similar peaks in most panels. The most pronounced differences appear in the NMP panel, where RMF-NL shows a distinctly broader distribution shifted toward lower log-likelihood values compared to the tighter distributions of DDH, DDB, and GDFMX. For the mass-radius constraints of PSR J0437+4715, all four models show relatively similar distributions with comparable peaks. The NICER constraints for PSR J0030+0451 and PSR J0740+6620 show excellent agreement among all models, with nearly overlapping distributions. In the GW170817 panel, the models are virtually indistinguishable, demonstrating consistent performance for tidal deformability constraints. In the combined astrophysical likelihood panel (bottom right), all models show comparable peaks centered around log($\mathcal{L}^{\text{Astro}}$) $\approx -9$, with RMF-NL displaying a slightly broader distribution, though the differences remain modest.
\begin{figure}
    \centering
    \includegraphics[width=1.0\linewidth]{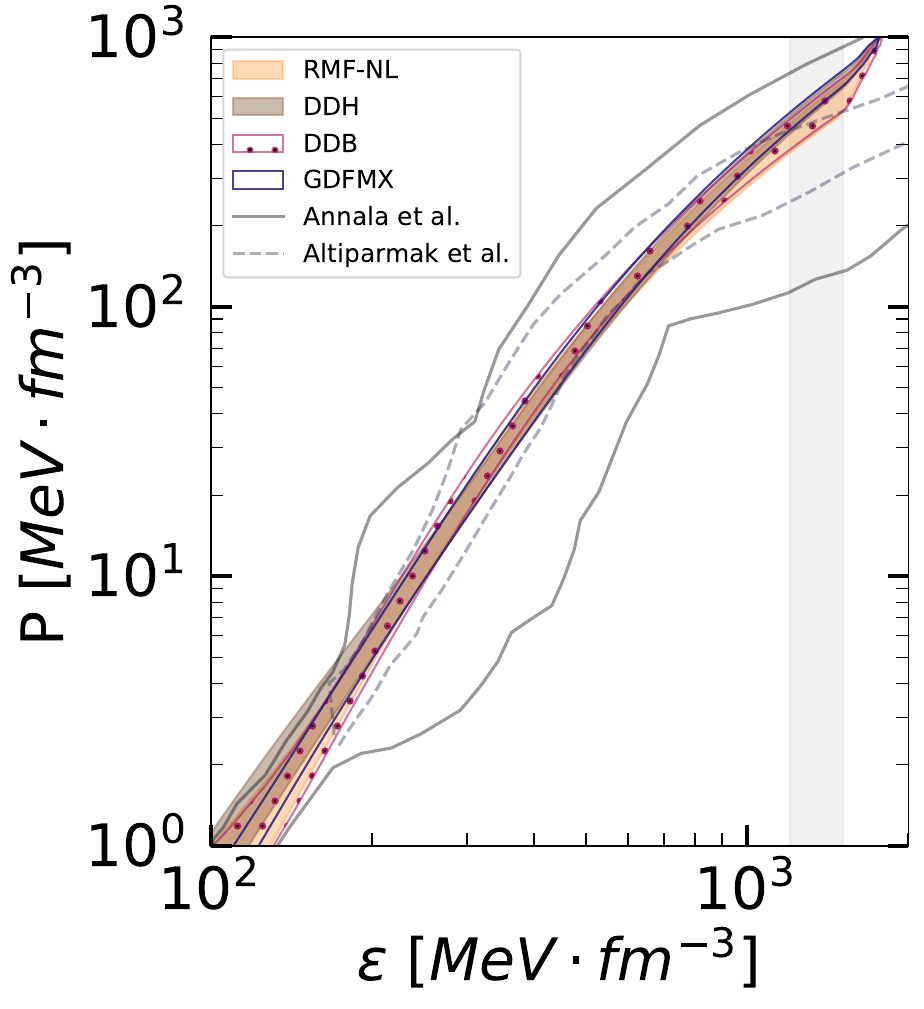}
    \caption{The 90\% CI of pressure versus energy density for phenomenological CEDF models evaluated in this study. The grey contour encloses all the EOS that satisfy the $\chi$EFT at low densities as in \citep{Hebeler2013} and several astronomical constraints \citep{Annala:2021gom}. The dashed grey line defines the region for which the PDF $\ge0.08$ \citep{Altiparmak:2022bke}, for the same set constraints. The light grey band defines the central baryonic density of the maximum mass configurations, above which no constraints have been imposed.}
    \label{fig:EOS}
\end{figure}

Recent advances in chiral Effective Field Theory ($\chi$EFT)  have enabled precise determinations of the pure neutron matter (PNM) equation of state (EOS), providing critical constraints for modeling dense matter within neutron stars. These calculations are valid for densities up to $\sim 2 \times n_0$, where $n_0=0.16$ fm$^{-3}$. The recent Ref. \citep{Huth:2021bsp} compiled $\chi$EFT constraints from various sources \citep{Hebeler2013, Tews:2012fj, Lynn:2015jua, Drischler:2017wtt, Drischler:2020hwi, Gezerlis:2009iw}. Given that the binding energy ($E/A - m$, where $m$ represents the nucleon mass) serves as the fundamental quantity in $\chi$EFT calculations, and the pressure is derived from it, we opt for the overlapping envelope of E/A for all these constraints in this study, as in \citep{Huth:2021bsp}. 
Figure ~\ref{fig:EOS_pnm} presents the 90\% credible intervals for the energy per nucleon of pure neutron matter (subtracting the nucleon rest mass), $E/A - m$, as a function of the baryon density $\rho$ across two panels. The left panel shows the baseline CEDF models (RMF-NL, DDH, DDB, and GDFMX). We also show chiral effective field theory ($\chi$EFT) bands (labeled “$\chi\mathrm{EFT}$”). In general,  the EOS predictions from RMF-NL, DDB, and DDH show a similar distribution inside the $\chi$EFT band. The GDFMX tends to show reduced values at low densities near the bottom of the band and increased values at high densities toward the top of the band (the GDFMX band is limited by thick lines). The right panel includes density-dependent variants (DDBy, DDHy) alongside DDB and DDH. The density-dependent models labeled `model'$y$ include an extra parameter in the definition of the coupling $\Gamma_\varrho$ as defined in Eq.~(\ref{y}). It can be seen from the figure that both models with and without the ``$y$'' parameter show similar results for the PNM EOS, as the ``$y$'' parameter primarily affects the high-density regime. The detailed effects on various EOS properties will be discussed later in the article.

\begin{table*}[t]
\caption{Nuclear Matter Properties:  The mean, minimum and maximum values at 90\% CI are given for the binding energy per nucleon $\epsilon_0$, the incompressibility $K_0$, the symmetry energy $J_{\rm sym, 0}$ and its slope $L_{\rm sym, 0}$ and curvature $K_{\rm sym, 0}$, defined at saturation density}
\centering
\begin{tabular}{lccccc}
\hline
\textbf{Model} & \textbf{$\epsilon_0$} & \textbf{$K_0$} & \textbf{$J_{\rm sym, 0}$} & \textbf{$L_{\rm sym, 0}$} & \textbf{$K_{\rm sym, 0}$} \\ \hline
DDH (\citep{Typel1999}) & $-15.99_{-0.31}^{+0.31}$ & $243_{-48}^{+43}$ & $32.0_{-2.6}^{+2.5}$ & $56.8_{-14.7}^{+17.3}$ & $-112_{-23}^{+36}$ \\
DDHy & $-15.99_{-0.30}^{+0.31}$ & $243_{-45}^{+44}$ & $32.2_{-2.6}^{+2.5}$ & $62.4_{-15.3}^{+17.3}$ & \; \: $-76.5_{-42.8}^{+49.9}$ \\
DDB (\citep{2022ApJ...930...17M}) & $-16.00_{-0.32}^{+0.31}$ & $227_{-30}^{+40}$ & $31.5_{-2.8}^{+2.6}$ & $42.6_{-12.7}^{+13.8}$ & $-105_{-34}^{+45}$ \\
DDBy & $-16.00_{-0.32}^{+0.32}$ & $221_{-29}^{+40}$ & $31.9_{-2.8}^{+2.6}$ & $48.7_{-13.8}^{+14.3}$ & \; \: $-65.2_{-55.8}^{+57.8}$ \\
RMF-NL (\citep{Mueller1996}) & $-16.00_{-0.31}^{+0.31}$ & $253_{-37}^{+39}$ & $31.4_{-2.6}^{+2.6}$ & $43.4_{-11.4}^{+14.2}$ & $-145_{-44}^{+67}$ \\
GDFMX (\citep{Gogelein:2007qa}) & $-16.00_{-0.32}^{+0.33}$ & $229_{-53}^{+51}$ & $32.4_{-2.4}^{+2.2}$ & $69.4_{-17.3}^{+16.9}$ & \; \: $-3.96_{-81.48}^{+75.41}$ \\
\hline
\end{tabular}
\label{tab:NMP}
\end{table*}

Fig. \ref{fig:EOS} shows the 90\% credible interval (CI) of the pressure versus the energy density posterior for all CEDF models considered. The grey envelope was obtained from Ref. \citep{Annala:2021gom,Altiparmak:2022bke}, defined by applying constraints from $\chi$EFT, pQCD and  multi-messenger astrophysical observations to the equation of state (EOS) of ultra-dense matter in neutron stars (NSs).  By analyzing a large ensemble of randomly generated EOSs, using a parametrized piecewise-linear speed-of-sound approach, that satisfy theoretical stability and causality requirements, the authors of \citep{Annala:2021gom} derive robust bounds on NS properties such as maximum mass, spin, compactness, and tidal deformability. The envelope is bounded at low densities by $\chi$EFT and at extremely high densities by pQCD computations. Besides they have considered the following astronomical constraints: i) a maximum mass $M_{TOV}>2 M_\odot$ to take into account the mass of the pulsars J0348+0432 \citep{Antoniadis2013} and PSR J0740+6620 \citep{Fonseca_2021}, ii) the  NICER  constraints for J0740+6620 and J0030+0451 by imposing $R(2M_\odot)> 10.75$~km and  $R(1.1 M_\odot)>10.8$~km, iii) the detection of GW170817 by  LIGO/Virgo by imposing $\tilde \Lambda<720$. The gray dashed lines show the dense PDF ($\ge 0.08$) obtained in ~\citep{Altiparmak:2022bke}, featuring a continuous sound speed imposing the same constraints. Notably, {the CEDF models are more constrained than the model-agnostic} description of the EOS. All our CEDF models fall within the highest probability region derived in \citep{Altiparmak:2022bke}
{except at the largest densities close to the maximum mass central densities. This reflects the fact that no phase transition to quark matter has been considered in our data sets, and our description does not allow the low-density $\chi$EFT EOS to be connected to the high-density pQCD EOS. However, note that the band defined by our datasets falls inside the grey envelope, as it should since the datasets have been constrained by the same ab-initio calculations and similar astronomical constraints. }. 

{Table~\ref{tab:NMP} summarizes the main nuclear matter properties of the six data sets. The mean and minimum and maximum values at 90\% CI are given for the binding energy per nucleon, the incompressibility, the symmetry energy and its slope and curvature, where all quantities are defined at saturation density. GDFMX shows the broadest range for incompressibility [176, 280]~MeV, while DDB predicts values in the range [197, 267]~MeV. DDH and RMF-NL give comparable ranges of [195, 286]~MeV and [216, 292]~MeV, respectively, with RMF-NL showing the highest central value of 253~MeV.}
{These values fall within the range discussed in the literature \citep{Huth:2020ozf,Stone:2014wza}.} Adding the parameter $y$ does not affect much the isoscalar properties, as expected. All models predict similar values for the symmetry energy at saturation. The parameter $y$ increases this value by $\sim 0.2-0.4$~MeV.  For all models, the 90\% CI corresponds to a range of $\sim \pm 2.5$~MeV. It is the slope and the curvature of the symmetry energy that most distinguish the models: for the slope, GDFMX gives the highest values, with a median of 70~MeV; DDB {and RMF-NL} give the lowest slope {$\sim43$}~MeV and DDH is {$\gtrsim30$}\% larger. The parameter $y$ increases the median by {$\sim 6$}~MeV. {However, all values fall within the range obtained by analyzing experimental data, ab-initio calculation or astrophysical observations discussed \cite{Tsang:2012se,Centelles:2008vu,Lattimer2013,oertel2017equations}.} {The symmetry energy curvature is a quantity that is much less constrained than the two previous properties. In \citep{Li:2021thg}, the authors  have obtained $K_{\text{sym}}=-107\pm 88$MeV at 68\% CI from analysis of 16 neutron star observables. However, note that including the scalar isovector meson and/or tensor interactions in CEDF, values as large as $\sim 500$ MeV were obtained when trying to reconcile results from the Lead Radius Experiment (PREX) and Calcium Radius
Experiment (CREX) Collaborations \citep{Reed:2023cap,Salinas:2023qic}.} While for the previous properties at 90\% CI the spread of values is similar for all datasets, for $K_{\text{sym}}$ RMF-NL and DDB show the smallest values, allowing values below -100 MeV at 90\% CI and  GDFMX shows a positive value for the median, although close to zero. 
Including the $y$ parameter pushes $K_{\text{sym}}$ to larger values and DDBy comes close to zero. The isovector properties have direct effect on density dependence of the  proton fraction that is discussed in one of the next subsections.

We have identified a notable prior dependency in the DDH/DDHy model, particularly for the $d_\sigma$ parameter, which significantly affects the nuclear matter incompressibility $K_0$ posterior distribution. Through systematic tests varying the prior range of $d_\sigma$ from 0 to 2, we observed substantial changes in the total likelihood distributions (Figure \ref{fig:ddh_prior_fix}). As the prior range narrows, the likelihood improves noticeably, though changes become negligible below $d_\sigma \in [0, 0.5]$. The impact on $K_0$ is particularly striking: the central value varies from approximately 200~MeV for $d_\sigma \in [0, 0.5]$ to 280~MeV for $d_\sigma \in [0, 2.0]$. For our final analysis, we adopted $d_\sigma \in [0, 1]$, yielding $K_0 = 243_{-48}^{+43}$~MeV at 90\% CI for the DDH model (Table~ \ref{tab:NMP}). This result agrees well with \citep{Li:2025oxi} (scenario F1) who obtained $K_0 = 244.8_{-33.7}^{+36.4}$~MeV (at 68\% CI), where $d_\sigma$ prior is taken in the range 0 to 1.5. It is important to note that in the DDH model, the parameters ($b_\sigma$ and  $c_\sigma$) are calculated from the priors and are inversely proportional to $d_\sigma$, meaning narrower priors on $d_\sigma$ result in wider ranges for these dependent parameters.

While one might expect the Bayesian sampler to naturally converge to the optimal parameter region regardless of prior width, this behavior suggests either a limitation in the sampling efficiency or a genuine multi-modal structure in the posterior. It should be emphasized that these findings are specific to our current likelihood setup, which includes constraints from nuclear matter properties, NICER observations (J0030, J0437, J0740), GW170817, chiral EFT for pure neutron matter, {and pQCD at 1.2 fm$^{-3}$}. Adding or removing different constraints may alter this scenario. This prior sensitivity analysis highlights the importance of carefully examining prior dependencies in complex nuclear physics models, particularly when multiple parameters are interdependent through theoretical relationships. Future work should investigate whether alternative parameterizations or sampling strategies could mitigate this prior dependency while maintaining physical consistency.

\begin{figure*}
    \centering
    \includegraphics[width=1.0\linewidth]{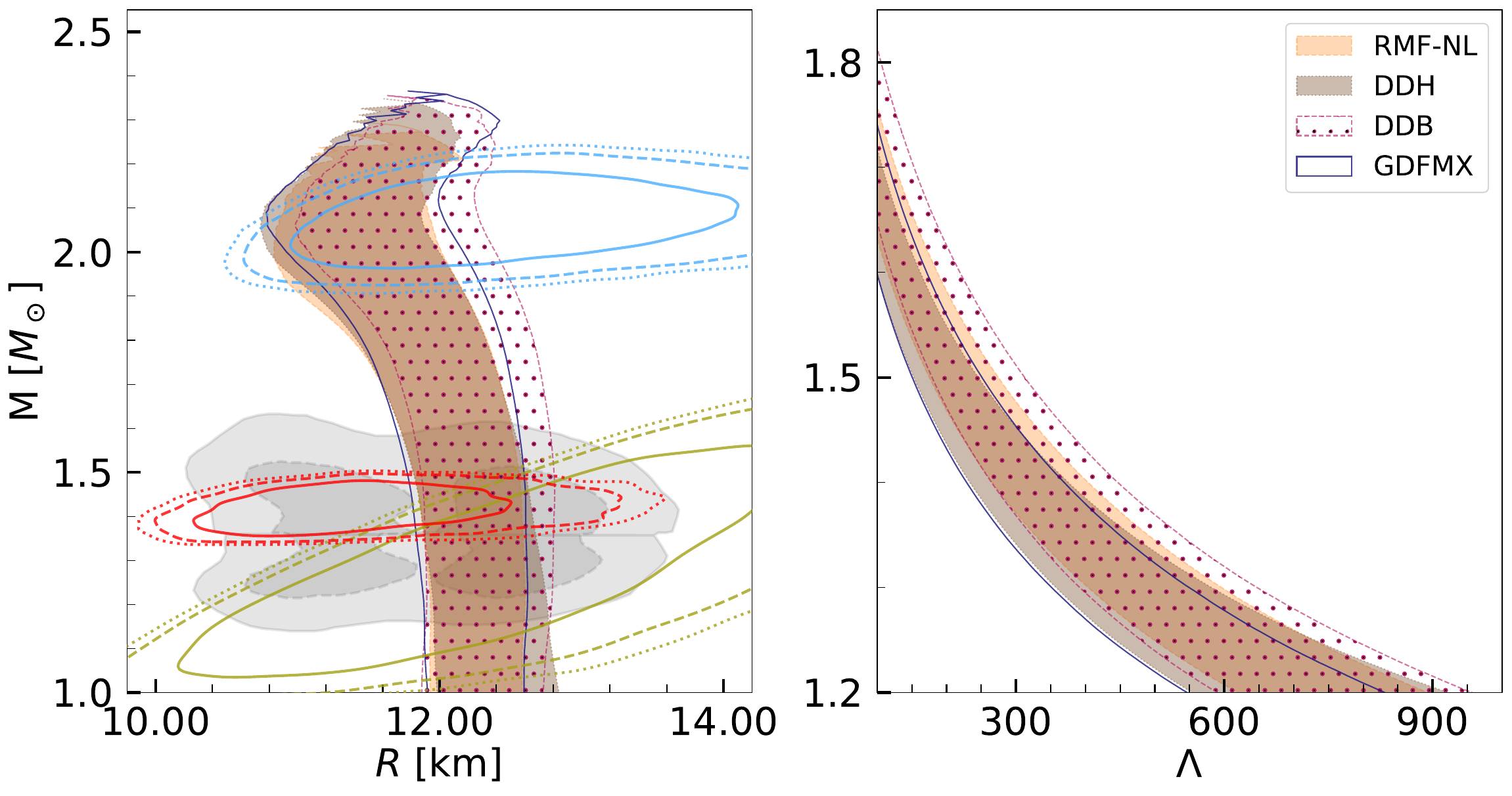}
    \caption{The 90\% credible interval (CI) posterior distribution of the neutron star's mass-radius $P (R|M )$ (left side) and mass-tidal deformability $P (\Lambda|M )$ (right side) is derived from different CEDF models.   We also compare the 1, 2 and 3 $\sigma$ (full, dashed and dotted lines respectively) CI for the two-dimensional posterior distributions within the mass-radius parameter space of the millisecond pulsar PSR J0030+0451, shown with olive green lines  \citep{Vinciguerra_2024} alongside with PSR J0740+6620, depicted with light blue lines \citep{salmi2024} and  PSR J0437+4715 with red lines \citep{Choudhury:2024xbk}, all derived from NICER X-ray observations. Additionally, the gray region represent the EOS independent MR posterior derived from GW170817 tidal deformability measurement \citep{LIGOScientific:2018cki}}
    \label{fig:MRT}
\end{figure*}

\subsection{Astrophysical predictions \label{subsec:mr}}

{Several studies have been carried out to constrain phenomenological hadronic models using Bayesian inference to impose astronomical and nuclear matter constraints. These include several CEDF descriptions for RMF-NL \citep{Traversi:2020aaa,Zhu:2022ibs,Pradhan:2022txg,Malik:2023mnx,Providencia:2023rxc,2024MNRAS.529.4650H,Imam:2024gfh,Passarella:2025zqb}, CEDF with  density dependent  couplings, as in DDB parametrization \citep{2022ApJ...930...17M,Beznogov:2022rri},
DDH parametrization \citep{Providencia:2023rxc,Char:2025zdy,Li:2025oxi}, GFDM parametrization \citep{Char:2023fue,Scurto:2024ekq,Char:2025zdy}, including the $\delta$-meson \citep{Santos:2024aii,Scurto:2025bqr}, within mean field models with chiral symmetry \citep{Malik:2024qjw,Marquez:2024bzj}, CEDF models including hyperons \citep{Li:2022cfd,Malik:2022jqc,2025MNRAS.536.3262H}, as well as non-relativistic models with Skyrme forces \citep{Beznogov:2023jqp,Imam:2024gfh,Beznogov:2024vcv}. The probability distributions obtained for the mass-radius curves depend not only on the model, but also on the prior and the constraints imposed to determine the parameters of the model. Therefore, it is not easy to compare the different results, and in the following we consider the four models we have introduced in Sec. \ref{sec:models} under the same constraints.}

Figure \ref{fig:MRT} presents a detailed comparison of the properties of neutron stars using observational and theoretical approaches. The left panel illustrates the mass-radius relationship, while the right panel focuses on mass-tidal deformability. Both panels display posterior distributions at a 90\% credible interval derived from four CEDF models: RMF-NL (orange shading), DDH (tan shading), DDB (red dotted region), and GDFMX (blue outlined region). Observational constraints are overlaid for comparison: 
the left panel includes 1, 2 and 3 $\sigma$  credible intervals (CI) from NICER X-ray observations of PSR J0030+0451, PSR J0740+6620 and  PSR J0437+4715 depicted in olive green, light blue, and red lines, based on analyses by \citet{Vinciguerra_2024},  \citet{salmi2024} and \citet{Choudhury:2024xbk}. 
{In addition, we show the 50\% and 90\% CI of the  mass radius posterior derived from the tidal deformability measurement of GW170817, based on analyses by \citep{LIGOScientific:2018cki}}. 

{The right panel shows the 90\% CI of the dimensionless tidal deformability for all models. It remains below 600 for a neutron star with a mass of 1.4 $M_\odot$. This result is consistent with constraints derived from gravitational wave observations of the binary neutron star merger event GW170817 \citep{LIGOScientific:2018cki}. A recent analysis in a Bayesian statistical framework has combined constraints from laboratory nuclear measurements with chiral effective field theory to predict neutron star properties. Their analysis found that the tidal deformability of a 1.4 solar mass neutron star falls within a 95\% credible interval of 136$<\Lambda<$ 519 \citep{Lim:2018bkq}, which is consistent with the upper limit from GW170817 observations. They also showed a strong correlation between the tidal deformability and the pressure of neutron star matter at twice the nuclear saturation density.} {The results presented in Figure \ref{fig:cor2rho0} and Table \ref{tab:cor2rho0} corroborate this relationship, showing that the correlation between $\Lambda_{1.4}$ and pressure is maximized at $\rho \approx 1.5 - 2~ n_0$ (i.e, 0.24 - 0.32 fm$^{-3}$) across all EOS models, with coefficients consistently exceeding 0.9 at this density while decreasing at higher densities.} The four models exhibit broad consistency with observational data across all mass ranges. For low-mass neutron stars, all models predict comparable radii, with $R_{1.4}$ values clustered around 12.2-12.9 km, and similarly comparable tidal deformabilities at 1.4 $M_\odot$.

\begin{table*}[!t]
\caption{Median values and 90\% CI for key neutron star properties of phenomenological CEDF models (RMF-NL, DDH, DDHy, DDB, DDBy, GDFMX) evaluated in this study. Parameters include maximum neutron star mass ($M_{\text{max}}$ in $M_\odot$), maximum central energy density ($\epsilon_{\rm c, max}$ in MeV fm$^{-3}$), radius for maximum mass $R_{\rm max}$ in km, number density for maximum mass $\rho_{\rm max}$ in fm$^{-3}$, proton fraction at $\rho_{\rm max}$ as $X_{\rm p, max}$,  and neutron star radii ($R_{1.4}$, $R_{1.6}$, $R_{1.8}$ in km) at different masses, as well as the dimensionless tidal deformability $\Lambda$ for the same NS masses.}
\setlength{\tabcolsep}{6.5pt}
\renewcommand{\arraystretch}{1.15}
\begin{center}
\begin{tabular}{ccccccccccccc}
\hline \hline
                          &                         & \multicolumn{3}{c}{RMF-NL}                              &  & \multicolumn{3}{c}{DDH}                                &  & \multicolumn{3}{c}{DDHy}                               \\ \hline
\multirow{2}{*}{Quantity} & \multirow{2}{*}{Units}  & \multirow{2}{*}{Median} & \multicolumn{2}{c}{90 \% CI} &  & \multirow{2}{*}{Median} & \multicolumn{2}{c}{90 \% CI} &  & \multirow{2}{*}{Median} & \multicolumn{2}{c}{90 \% CI} \\
                          &                         &                         & Min           & Max          &  &                         & Min           & Max          &  &                         & Min           & Max          \\ \hline
$M_{\rm max}$             & $M_\odot$               & 2.029                   & 1.925         & 2.137        &  & 2.091                   & 1.995         & 2.177        &  & 2.078                   & 1.975         & 2.159        \\
$\epsilon_{\rm c, max}$   & MeV fm$^{-3}$           & 1426                    & 1274          & 1509         &  & 1426                    & 1348          & 1509         &  & 1426                    & 1348          & 1509         \\
$R_{\rm max}$             & km                      & 10.70                   & 10.37         & 11.15        &  & 10.62                   & 10.27         & 10.94        &  & 10.65                   & 10.30         & 10.97        \\
$\rho_{\rm c, max}$       & fm$^{-3}$               & 1.112                   & 1.020         & 1.187        &  & 1.102                   & 1.038         & 1.167        &  & 1.102                   & 1.040         & 1.169        \\
$X_{\rm p, max}$          & ...                     & 0.127                   & 0.112         & 0.142        &  & 0.128                   & 0.125         & 0.143        &  & 0.174                   & 0.130         & 0.249        \\
$^{*} \rho_{\rm Urca}$    & fm$^{-3}$               & 1.010                   & 0.686         & 1.190        &  & 0.460                   & 0.320         & 0.690        &  & 0.520                   & 0.350         & 1.030        \\
$^{*} M_{\rm Urca}$       & $M_\odot$               & 2.031                   & 1.884         & 2.143        &  & 1.376                   & 0.956         & 1.879        &  & 1.565                   & 1.026         & 2.091        \\
$R_{1.4}$                 & \multirow{4}{*}{km}     & 12.21                   & 11.87         & 12.58        &  & 12.26                   & 11.87         & 12.65        &  & 12.41                   & 12.01         & 12.86        \\
$R_{1.6}$                 &                         & 12.10                   & 11.73         & 12.50        &  & 12.12                   & 11.72         & 12.52        &  & 12.25                   & 11.82         & 12.69        \\
$R_{1.8}$                 &                         & 11.86                   & 11.39         & 12.33        &  & 11.90                   & 11.42         & 12.33        &  & 11.98                   & 11.47         & 12.44        \\
$R_{2.0}$                 &                         & 11.03                   & 10.05         & 12.01        &  & 11.44                   & 10.01         & 12.00        &  & 11.47                   & 9.96          & 12.04        \\
$\Lambda_{1.4}$           & \multirow{4}{*}{\ldots} & 327                     & 266           & 404          &  & 295                     & 234           & 375          &  & 328                     & 256           & 442          \\
$\Lambda_{1.6}$           &                         & 146                     & 115           & 187          &  & 128                     & 100           & 162          &  & 138                     & 106           & 182          \\
$\Lambda_{1.8}$           &                         & 62                      & 46            & 85           &  & 54                      & 39            & 71           &  & 57                      & 40            & 76           \\
$\Lambda_{2.0}$           &                         & 18                      & 6             & 34           &  & 20                      & 5             & 30           &  & 20                      & 5             & 30           \\ \hline
\end{tabular}
\begin{tabular}{ccccccccccccc}
\hline
                          &                         & \multicolumn{3}{c}{DDB}                                &  & \multicolumn{3}{c}{DDBy}                               &  & \multicolumn{3}{c}{GDFMX}                              \\ \hline
\multirow{2}{*}{Quantity} & \multirow{2}{*}{Units}  & \multirow{2}{*}{Median} & \multicolumn{2}{c}{90 \% CI} &  & \multirow{2}{*}{Median} & \multicolumn{2}{c}{90 \% CI} &  & \multirow{2}{*}{Median} & \multicolumn{2}{c}{90 \% CI} \\
                          &                         &                         & Min           & Max          &  &                         & Min           & Max          &  &                         & Min           & Max          \\ \hline
$M_{\rm max}$             & $M_\odot$               & 2.097                   & 1.981         & 2.196        &  & 2.089                   & 1.975         & 2.187        &  & 2.122                   & 2.032         & 2.209        \\
$\epsilon_{\rm c, max}$   & MeV fm$^{-3}$           & 1348                    & 1204          & 1426         &  & 1348                    & 1204          & 1426         &  & 1426                    & 1274          & 1509         \\
$R_{\rm max}$             & km                      & 10.92                   & 10.53         & 11.44        &  & 10.95                   & 10.55         & 11.47        &  & 10.66                   & 10.32         & 11.12        \\
$\rho_{\rm c, max}$       & fm$^{-3}$               & 1.050                   & 0.957         & 1.127        &  & 1.051                   & 0.956         & 1.129        &  & 1.093                   & 1.008         & 1.166        \\
$X_{\rm p, max}$          & \ldots                  & 0.122                   & 0.119         & 0.125        &  & 0.154                   & 0.123         & 0.218        &  & 0.205                   & 0.154         & 0.254        \\
$^{*} \rho_{\rm Urca}$    & fm$^{-3}$               & \ldots                  & \ldots        & \ldots       &  & 0.630                   & 0.410         & 1.100        &  & 0.380                   & 0.300         & 0.600        \\
$^{*} M_{\rm Urca}$       & $M_\odot$               & \ldots                  & \ldots        & \ldots       &  & 1.875                   & 1.345         & 2.140        &  & 1.047                   & 0.706         & 1.799        \\
$R_{1.4}$                 & \multirow{4}{*}{km}     & 12.32                   & 11.89         & 12.81        &  & 12.46                   & 12.01         & 12.95        &  & 12.20                   & 11.82         & 12.61        \\
$R_{1.6}$                 &                         & 12.27                   & 11.81         & 12.78        &  & 12.39                   & 11.90         & 12.89        &  & 12.11                   & 11.69         & 12.56        \\
$R_{1.8}$                 &                         & 12.13                   & 11.60         & 12.66        &  & 12.21                   & 11.64         & 12.76        &  & 11.93                   & 11.46         & 12.44        \\
$R_{2.0}$                 &                         & 11.70                   & 10.19         & 12.39        &  & 11.72                   & 10.17         & 12.45        &  & 11.56                   & 10.33         & 12.17        \\
$\Lambda_{1.4}$           & \multirow{4}{*}{\ldots} & 335                     & 265           & 447          &  & 365                     & 283           & 483          &  & 288                     & 229           & 369          \\
$\Lambda_{1.6}$           &                         & 159                     & 121           & 219          &  & 168                     & 127           & 232          &  & 129                     & 99            & 171          \\
$\Lambda_{1.8}$           &                         & 74                      & 53            & 106          &  & 76                      & 53            & 110          &  & 57                      & 42            & 80           \\
$\Lambda_{2.0}$           &                         & 28                      & 7             & 46           &  & 28                      & 6             & 46           &  & 23                      & 9             & 35           \\ \hline
\end{tabular}
\end{center}
\footnotesize
\begin{tablenotes}
\item \hspace{0.8cm} \parbox{0.85\linewidth}{$^{*}$ Not all models in each set exhibit Urca cooling. Among the models, the percentage with Urca threshold proton fractions is: RMF-NL (9\%), DDH (7\%), DDHy (82\%), DDB (0\%), DDBy (69\%), GDFMX (99\%).}
\end{tablenotes}
\label{tab:NSprop}
\end{table*}

The noticeable differences appear only at higher masses, where variations in stiffness become significant. The overlap between model predictions and observational bounds shows that each CEDF model aligns with current constraints from X-ray. The DDB and GDFMX allow for more massive stars $\sim$2.4$M_\odot$, while RMF-NL predicts the smallest maximum masses, {close to 2.3$M_\odot$}. Table \ref{tab:NSprop} summarizes some of the most important properties of NSs.  Although at 90\% CI the maximum predicted masses lie between 2.14 and 2.21$M_\odot$, every model can represent neutron stars having masses upwards of approximately 2.3$M_\odot$. Comparing the behavior of the different models with respect to the radius of 1.4 and 2.0$M_\odot$ stars (respectively, $R_{1.4}$ and $R_{2.0}$), we conclude that most models predict a median $R_{2.0}$ above 11.4~km, with DDH/DDHy predicting values around 11.45~km, DDB/DDBy around 11.7~km, and GDFMX at 11.56~km, while RMF-NL shows the lowest median value at 11.03~km. {The minimum value of $R_{2.0}$ across all models at the 90\% credible interval ranges from 9.96 to 10.33~km.} The minimum radius of PSR J0740 was found to be above 11.1~km at 95\% CI \citep{Riley:2021pdl,Miller2021,salmi2024}. For $R_{1.4}$, the median is between 12.2 and 12.5 km, with the minimum above 11.8 km and the maximum below 13 km. These results are also broadly consistent with the EOS-independent analysis presented in \citep{Huang:2024wig}, underscoring the robustness of different models in predicting the neutron star radius at 1.4 $M_{\odot}$, $R_{1.4}$. Models with the $y$ parameter predict radii about 150 m larger. Note that the values obtained for $R_{1.6 ~M_\odot}$ satisfy the constraints obtained from the analysis of GW170817, considering that there was no prompt collapse, indicating that the radius $R_{1.6 ~M_\odot}$ should be larger than 10.68$^{+0.15}_{-0.04}$ km \citep{Bauswein:2017vtn}: at 90\% CI this constraint is satisfied by all datasets.

\bigskip

\begin{figure}[ht!]
    \centering
    \includegraphics[width=1.0\linewidth]{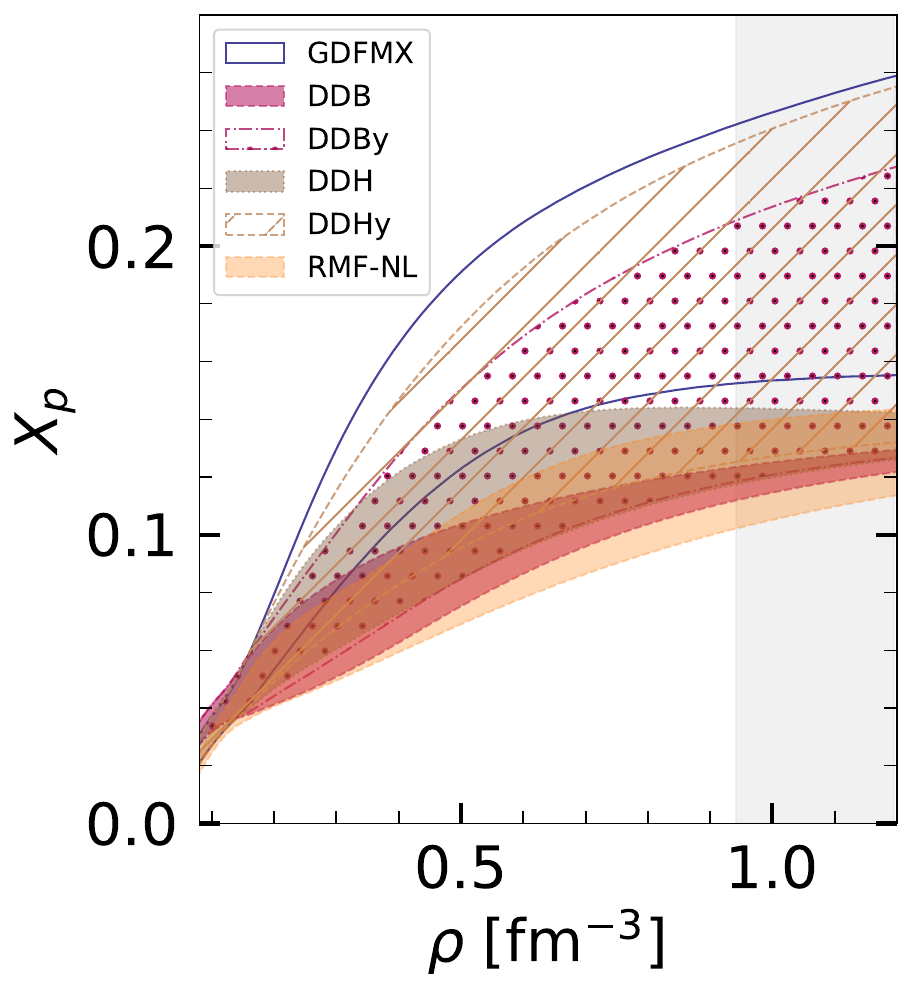}
    \caption{The posterior distribution of the 90\% CI for the proton fraction and multiple CEDF models as a function of the baryon density.}
    \label{fig:xp}
\end{figure}

\subsection{Proton fraction \label{subsec:yp}}

\begin{figure*}[ht!]
    \centering
    \includegraphics[width=1.0\linewidth]{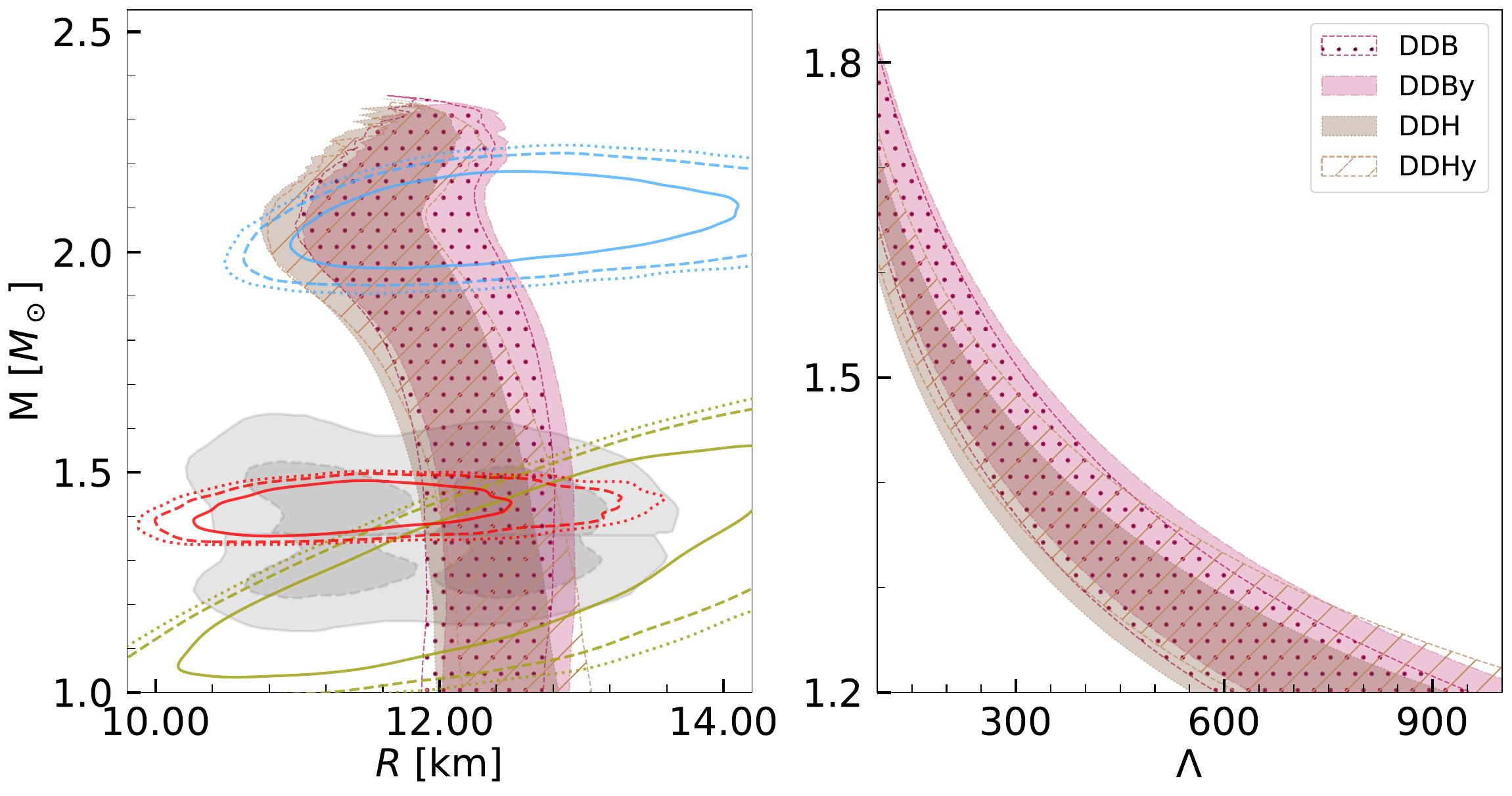}
    \caption{The 90\% CI posterior distribution of the neutron star's mass-radius $P (R|M )$ (left side) and mass-tidal deformability $P (\Lambda|M )$ (right side) is derived from the  models DDH and DDB with and without the $y$ parameter introduced in Eq. (\ref{y}). {We also compare the 1, 2 and 3 $\sigma$ (full, dashed and dotted lines respectively) CI for the two-dimensional posterior distributions within the mass-radius parameter space of the millisecond pulsar PSR J0030+0451, shown with green lines  \citep{Vinciguerra_2024} alongside with PSR J0740+6620, depicted with blue lines \citep{salmi2024} and  PSR J0437+4715 with red lines \citep{Choudhury:2024xbk}, all derived from NICER X-ray observations. Additionally, the gray region represents the EOS independent MR posterior derived from GW170817 tidal deformability measurement \citep{LIGOScientific:2018cki}}}
    \label{fig:MRT-ys}
\end{figure*}

\begin{figure*}
    \centering
    \includegraphics[width=1.0\linewidth]{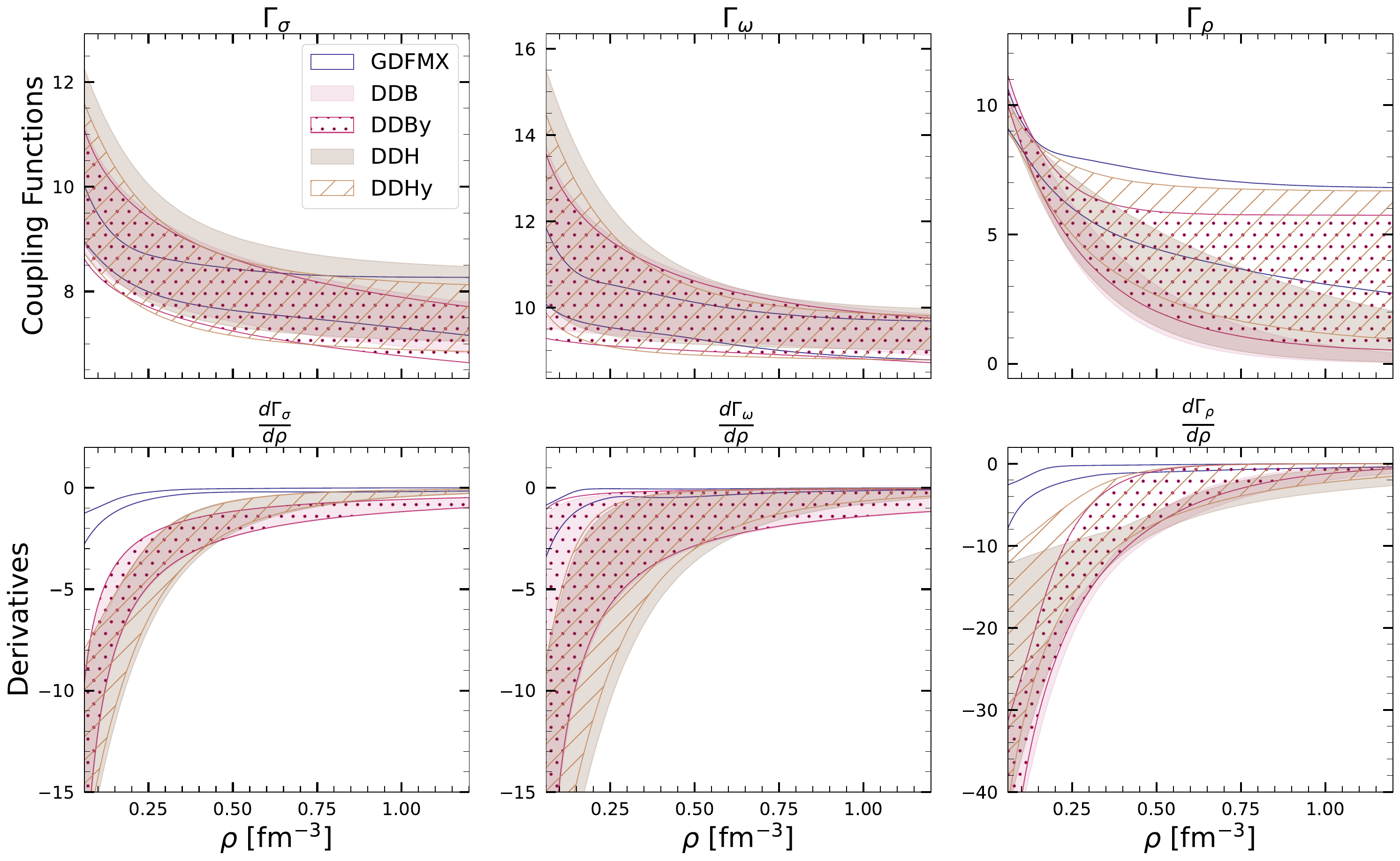}
    \caption{The 90\% CI variation of the coupling constants $\Gamma_\sigma$, $\Gamma_\omega$, and $\Gamma_\rho$ for the $\sigma$, $\omega$, and $\rho$ fields, along with their derivatives, as a function of number density $\rho$ across all density-dependent models employed in this study.}
    \label{fig:couplings}
\end{figure*}

One quantity that is interesting to compare is the predicted proton fraction inside NSs, since this dictates their behavior with respect to the possible opening of nucleonic direct Urca processes and is closely related to the dependence of the symmetry energy on the baryon density. 
{However, it has been shown that the extraction of the symmetry energy from the EOS of $\beta$-equilibrium matter comes associated with large uncertainties \cite{Imam:2021dbe,Mondal2021,Tovar2021,Essick2021}. Consequently, the composition of NS with respect to their proton content is poorly understood. The possibility of extracting NMP from the detection of tidal deformability has recently been discussed in \citep{Iacovelli:2023nbv}. In \citep{Marino:2024gpm},  an appropriate description of the NS cooling has shown to give important information about its composition. In particular,  the authors concluded that for young cold NSs, their EOS must be compatible with a fast cooling process, at least for some NS masses.  }

Figure \ref{fig:xp} shows the 90\% credible intervals for the proton fraction, \(X_p\), as a function of the baryon density, \(\rho\), derived from the posterior distributions of the different models for \(\beta\)-equilibrated neutron star matter. The DDB, DDH and RMF-NL show a quite small proton fraction at high densities (\(\rho > 0.5\,\text{fm}^{-3}\)).  In particular, the GDFMX model predicts a higher proton fraction at high densities than the other models. In addition, this model predicts proton fractions above 0.11 at quite low densities, favoring a fast cooling of low mass NS through the nucleonic direct Urca processes. If no muons had been included in the calculation, the opening of these processes would occur for $X_p\sim 0.11$, including muons the opening of these processes shift to a value of the proton fraction $0.11< X_p<0.148$. The band obtained for GDFMX at 90\% CI predicts that all stars with a baryonic density above 4$n_0$ in their interiors will suffer a fast cooling, if pairing is not considered \citep{Yakovlev:2000jp}. This narrow GDFMX band contrasts with the results of other works \citep{Char:2025zdy}. In principle, GDFMX can exhibit a wider range in the proton fraction than DDH and DDB  because the isovector sector of the functional includes more free parameters. However, the present study does not span the wide proton fraction band due to the different priors and constraints considered.

This point deserves some discussion. The coupling parameterization within GDFMX contains a $x^3$ term (see Eq. (\ref{gdfm})) that may give the origin to a nonphysical behavior of the EOS. In order to control this term, 
{as discussed in Sec. \ref{sec:models}, }
in our Bayes inference calculation we have imposed that the GDFMX couplings never increase with density, {in order to reproduce the behavior predicted within a DBHF calculation.} DDH and DDB parametrizations satisfy this condition automatically. As a result of this constraint, the coupling $d_i$ that multiplies the term $x^3$ in GDFMX is very small. The behavior of the proton fraction for this model is similar to that discussed in \citep{Scurto:2025bqr} where the term $x^3$ was not considered. To give more flexibility to the model the authors  have introduced the $\delta$-meson in the isovector channel. In \citep{Char:2025zdy}, the authors adapted a different strategy, {and, used an informed prior} that allows the occurrence of small and large proton fractions in the interior of NS.

Another aspect worth discussing is the behaviour of the DDB and DDH models at high densities. In these models, the coupling of the $\rho$-meson decreases exponentially with density, approaching zero at high density (see Eq. \ref{grho}). This behaviour reduces the symmetry energy, thereby favouring very asymmetric matter and preventing the onset of nucleon direct Urca processes in the stellar core. To address this limitation, a modified coupling of the $\rho$ meson was introduced in \citep{Malik:2022ilb}, incorporating an additional parameter y into the density dependence function, see Eq. (\ref{y}).
The parameter \(y\) may be tuned based on neutron star properties -- such as the stellar mass \(M_{dUrca}\) at which direct nucleon Urca processes become possible -- that are sensitive to the high-density behavior of the symmetry energy. Fig. \ref{fig:xp} shows that with the introduction of parameter \(y\), the DDHy and DDBy models exhibit an increased proton fraction compared to their base models (DDH and DDB). GDFMX shows the highest proton fraction, reflecting the  different parametrization of the \(\rho\)-meson coupling relative to the other density-dependent models. {The introduction of the $\delta$-meson as in the original model GDFM \citep{Gogelein:2007qa} and in \citep{Scurto:2024ekq}  will give more flexibility to the isovector channel. Note that the high density behavior of the symmetry energy is still an unknown, and phenomenological models need either experimental data or ab-initio calculations to be constrained.}

In Figure \ref{fig:MRT-ys}, we present a comparison of the posteriors of the mass-radius and mass-tidal deformability for the DDB and DDH base models with the newly introduced DDBy and DDHy models, which include the additional parameter $y$. The newly introduced parameter $y$ primarily influences the behavior of the proton fraction at high density, {see Fig. \ref{fig:xp}, allowing for large proton fractions inside NS contrary to the original models DDB and DDH. Observational data from NS cooling could constrain this parameter. In fact, its impact on NS properties as mass-radius and mass-tidal deformability curves is not strong: its main effect occurs for low-mass and intermediate-mass NS, predicting a larger radius and tidal deformability. This can be attributed to the stiffer symmetry energy associated with the models that include the $y$ parameter. As will be discussed in the following, where the Bayes evidence for these models is presented: the introduction of the additional parameter $y$ has only a negligible effect on the Bayes evidence, slightly increasing it (see Table \ref{tab:log_bayes}). This could indicate that the constraints imposed in the inference are not sensitive to the EOS isovector channel.}

In order to complete the discussion, we show in Fig. \ref{fig:couplings}, the density dependence of the couplings and their derivatives with respect to the baryonic density for all  models with density-dependent couplings discussed in the present study. The range of values spanned by each model dictates its flexibility to be able to describe different constraints simultaneously.

\subsection{Speed of sound \label{subsec:cs}}
\begin{figure}
    \centering
    \includegraphics[width=1.0\linewidth]{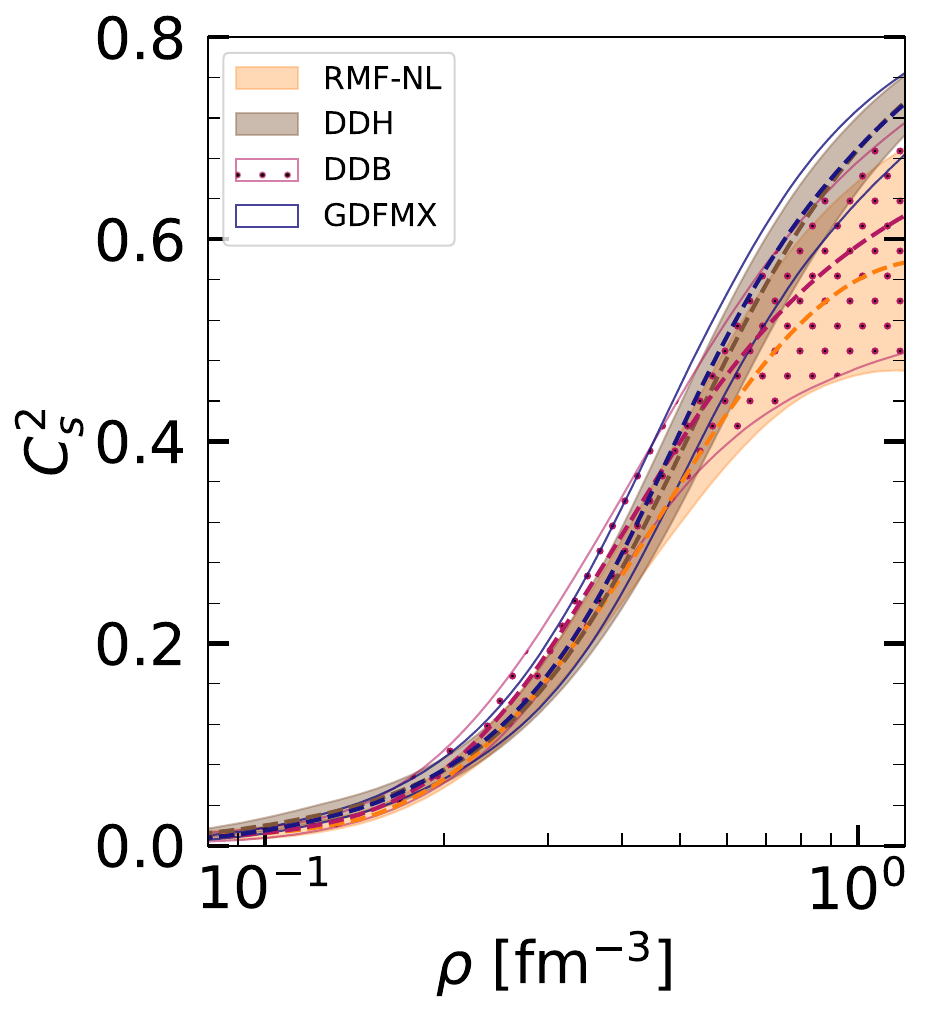}
    \caption{The 90\% CI for the square of the speed of sound, expressed as a function of baryon density, is derived from the posterior distribution for several phenomenological EOS models considered here.}
    \label{fig:cs2}
\end{figure}

Fig. \ref{fig:cs2}, illustrates the 90\% CIs for the square of the speed of sound, \( c_s^2 \), as a function of baryon density (\( \rho \)), derived from the posterior distributions of various phenomenological EOS models (RMF-NL, DDH, DDB, and GDFMX) considered in this work. All models show an increase in \( c_s^2 \) with baryon density, followed by a saturation at higher densities, consistent with theoretical expectations for dense matter in neutron stars.  The prediction of $c_s^2$ for GDFMX is lower at low density and higher at high density compared to other models, following a trend similar to that discussed in PNM earlier.

The GDFMX and DDH models show narrower credible intervals than the DDB and NL models, {and larger values at large densities, above 0.72 for GDFMX and 0.64 for DDH at 90\% CI for $\rho=1$fm$^{-3}$. For the other two models this lower limit is below 0.5.} The saturation behavior of \( c_s^2 \) reflects compliance with the causal limit (\( c_s^2 \leq 1 \)), a critical feature of physically consistent equations of state. The differences between models at higher densities underscore the sensitivity of \( c_s^2 \) to the assumptions underlying each CEDF parameterization, especially in the regime relevant to the extreme conditions of neutron star cores.  The behavior of the RMF-NL speed of sound at high densities was discussed in \citep{Malik:2023mnx}, where it was shown to be closely related to the term $\omega^4$ in the Lagrangian density of this model.

% \bigskip

\subsection{Bayes factor \label{subsec:bayesf}}

\begin{table}[ht!]
\caption{Log Bayes Evidence for the CEDF models studied, calculated using importance nested sampling in PyMultiNest. This approach provides significantly higher accuracy compared to conventional methods without importance sampling. DDHy and DDBy models include the $y$ parameter introduced in Eq. (\ref{y}).}
\centering
\setlength{\tabcolsep}{25.5pt}
\renewcommand{\arraystretch}{1.1}
\begin{tabular}{lc}
\hline \hline 
\textbf{Model} & \textbf{Log Bayes Evidence} \\ \hline
DDH  & $-48.23 \pm 0.09$ \\
DDHy  & $-48.17 \pm 0.07$ \\
DDB  & $-47.03 \pm 0.02$ \\
DDBy  & $-47.32 \pm 0.02$ \\
RMF-NL  & $-46.47 \pm 0.08$ \\
GDFMX  & $-49.45 \pm 0.50$ \\ \hline
\end{tabular}
\label{tab:log_bayes}
\end{table}

To evaluate model performance, we computed Bayes' evidence for each model, and the results are presented in table \ref{tab:log_bayes}. {The Bayes factor, defined as the ratio of marginal likelihoods between competing models, quantitatively measures the strength of evidence favoring one model over another. According to the Jeffreys' scale, differences in the logarithm of Bayes factors can be interpreted as: 0 to 1 ("minimal evidence"), 1 to 2.5 ("substantial evidence"), 2.5 to 5 ("strong evidence"), and $>5$ ("decisive evidence") \citep{kass}.} {Examining the Bayesian evidence reveals distinct groupings among the models. RMF-NL shows the highest evidence ($-46.47 \pm 0.08$), with DDB ($-47.03 \pm 0.02$) showing a minimal difference of 0.56 in log evidence, which falls within "minimal evidence" on Jeffreys' scale. DDBy ($-47.32 \pm 0.02$) shows a difference of 0.85, approaching the boundary between minimal and substantial evidence. In contrast, DDH ($-48.23 \pm 0.09$) and DDHy ($-48.17 \pm 0.07$) show differences of approximately 1.7-1.8 from the best model, indicating "substantial to strong evidence" against them, while GDFMX ($-49.45 \pm 0.50$) shows a difference of 3.0, constituting "strong evidence" against it. It is important to note that while Fig.~\ref{fig:loglike} compares only a subset of the constraints (NMP, PSR J0030+0451, PSR J0740+6620, PSR J0437+4715, and GW170817), the Bayes evidence was calculated using the entire data set. This comprehensive data set includes constraints from pQCD applied at 1.2~fm$^{-3}$ and $\chi$EFT constraints. In summary, considering the present set of constraints and priors, GDFMX is strongly disfavored, DDH and DDHy are substantially to strongly disfavored, while RMF-NL and DDB remain statistically most favored, with DDBy showing intermediate evidence. Future studies with improved observational data and refined priors will be essential for identifying the most plausible EOS candidate for neutron star matter, which may require, as discussed before, the generalization of the models considered.}

\section{Conclusion} \label{sec:conclusions}
In conclusion, our analysis using the \texttt{CompactObject} package has demonstrated that while all considered CEDF models broadly reproduce current astrophysical and theoretical constraints, subtle yet important differences persist among them. In particular, all models yield comparable posterior distributions across nuclear saturation properties, mass-radius relations, tidal deformability, and other EOS properties, {although DDH and GDFMX are disfavored when considering the Bayes factor, which also includes the $\chi$EFT and pQCD constraints}. These findings are further supported by the models’ behavior with respect to the pure neutron matter EOS, the evolution of the speed of sound and {proton fraction}  with each framework exhibiting distinct characteristics at supranuclear densities. The incorporation of the additional parameter in the DDHy and DDBy variants illustrates that adjustments in the high-density behavior of the symmetry energy can significantly influence the proton fraction without notably altering the macroscopic neutron star properties.

Figure~\ref{fig:prior_compare} compares the full prior domains (dashed black lines) with the 90\% credible intervals of posterior distributions (shaded regions) for all six CEDF models. The three columns show, from left to right: pure neutron matter energy per particle $(E/A - m)$ versus density with $\chi$EFT constraints, proton fraction $X_p$ in $\beta$-equilibrium, and neutron star mass-radius relations. The visualization demonstrates how Bayesian inference effectively constrains the broad priors to narrow posteriors consistent with all observational and theoretical data. It should be noted that for all these 6 CEDF model, the "prior" shown represents parameters already constrained by nuclear saturation properties (NMP) and the requirement $M_{\text{max}} > 1.7 M_\odot$, as the full unconstrained parameter space would yield unphysical nuclear matter properties.

Overall, the agreement between our phenomenological models and constraints from $\chi$EFT, pQCD, and multimessenger observations underscores the robustness of the selected approaches while also highlighting the sensitivity of dense matter predictions to the underlying EOS parameterizations. Future work with improved observational data and refined priors will be crucial to further constrain the EOS and pinpoint the most plausible description of neutron star matter. {A comparison of our results with other recent studies, in particular with \citep{Char:2025zdy} and \citep{Li:2025oxi}, has alerted us to the role of priors. Indeed, due to the nonlinearities of the models, the range of priors chosen can lead the inference to different regions of phase space.} 

{The present study identifies the main NS properties if matter only includes nucleons. In a next study, the inclusion of other degrees of freedom, such as hyperons or kaon condensates, should be considered in order to have a full picture of NS formed by hadronic matter. The missing input to these frameworks is the possible transition to a deconfined quark phase or a quarkionic phase. Note that we have restricted ourselves to a relativistic description in order to ensure causality. NS with a central density below 0.2 fm$^{-3}$ (twice saturation density) have masses below 0.5 (1.0) $M_\odot$ and it is expected that the minimum NS mass is on the order of one solar mass \citep{Suwa:2018uni,Strobel:2000mg,Lattimer:2000nx}.  We hope that a clear picture of hadronic NS properties will allow for the unambiguous identification of other types of matter. With respect to the model-agnostic approaches, the present work allows the identification of the regions in mass-radius space, among other possible physical quantities, that are not covered by an hadronic EOS. }

\section*{Acknowledgements}
\noindent This work was supported by the EURO HPC project (EHPC-DEV-2024D12-009) and FCT project (2024.12970.CPCA). The computational component of this research was carried out on the Deucalion HPC platform in Portugal, whose resources and technical assistance are gratefully acknowledged. This analysis required approximately 500,000 CPU hours to evaluate 30 million total samples across all models, with the GDFMX model accounting for approximately 40\% of the total computational cost due to its complex parameter space and coupling constraints. J. C., T. M. and C.P. received support from Fundação para a Ciência e a Tecnologia (FCT), I.P., Portugal, under the projects UIDB/04564/2020 (doi:10.54499/UIDB/04564/2020), UIDP/04564/2020 (doi:10.54499/UIDP/04564/2020). H.C. and S.S. acknowledges support from the Arts \& Sciences Fellowship of Washington University in St Louis. We acknowledge Micaela Oertel for the thorough review of the manuscript and constructive feedback on theoretical constraints and EoS formalism that significantly enhanced the quality of this work. We thank Laura Tolos for valuable discussions on dense matter EoS consistency and theoretical constraints, and Anna Watts for expert guidance on X-ray observational constraints and NICER data implementation. We also thank Prasanta Char for valuable cross-verification work on the GDFM model implementation.

\section*{Data Availability}
The full posterior distributions, derived from a sample of approximately {145,000} sample across all models considered in this study--averaging 24,000 samples per model--were obtained by evaluating {30} million total samples. These distributions encompass equation of state (EOS) data, nuclear saturation properties, and mass-radius-tidal deformability results for all models analyzed. This dataset is publicly available on Zenodo [\href{https://doi.org/10.5281/zenodo.17465229}{URL}] \citep{cartaxo_2025_15482081}. To enhance accessibility and streamline analysis, we have developed a dedicated graphical user interface (GUI) application, designed to assist researchers in extracting, filtering, and visualizing targeted subsets of data without the need for command-line proficiency. The GUI application is available at: \url{https://github.com/tuhinbits/CompactObject-CEDF-EOS-Database-GUI}. The application supports the selection of specific model types, filtering based on nuclear matter properties (e.g., incompressibility or symmetry energy), and the generation of customized EOS or mass-radius curves with user-defined credible intervals. It also offers advanced features such as wildcard-based database queries, simultaneous application of multiple filters, and export options in various formats. The Bayesian inference framework employed in this study, are openly accessible in the \texttt{CompactObject} repository, accompanied by detailed documentation. Zenodo repository of \texttt{CompactObject} package: \url{https://zenodo.org/records/14181695} for Version 2.0.0, Github: \url{https://github.com/ChunHuangPhy/CompactObject}, documentation: \url{https://chunhuangphy.github.io/CompactObject/}.

\appendix
\begin{figure}[htb!]
    \centering
    \includegraphics[width=1\linewidth]{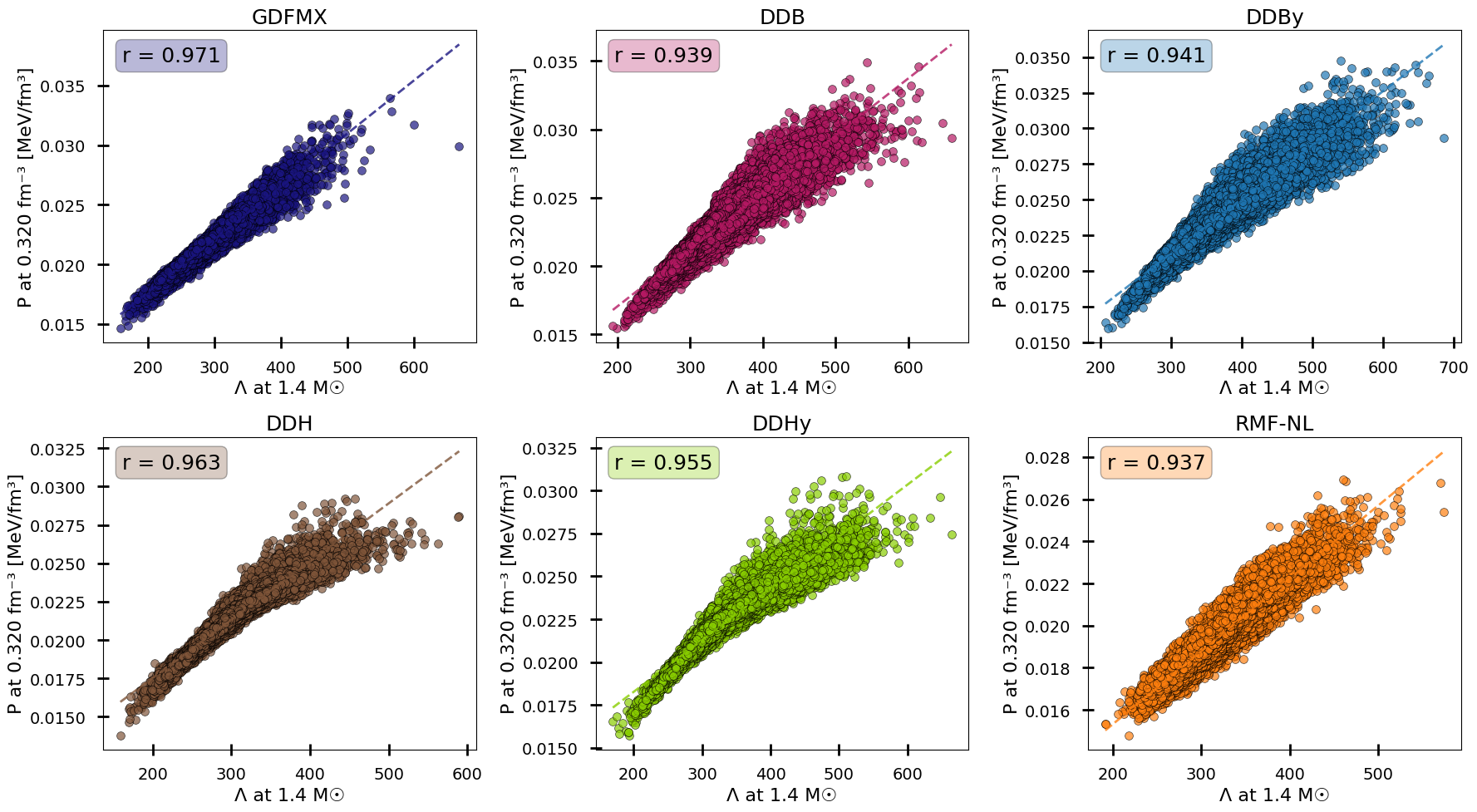}
    \caption{The Pearson correlation coefficient linking the pressure at 0.32 fm$^{-3}$ (approximately $2 n_0$) with the tidal deformability of a 1.4 M$_\odot$ neutron star is examined across all EOS sets in this study.}
    \label{fig:cor2rho0}
\end{figure}

\begin{table}[ht!]
\centering
\caption{Correlation coefficient $r$ between $\Lambda_M (M = 1.4\,M_\odot)$ and pressure at different densities $\rho$.}
\begin{tabular}{c|c|c|c|c}
\hline \hline 
\multirow{2}{*}{\textbf{Model}} & $\rho = 0.240$ fm$^{-3}$ & $\rho = 0.320$ fm$^{-3}$ & $\rho = 0.400$ fm$^{-3}$& $\rho = 0.480$ fm$^{-3}$ \\
& ($\sim$ 1.5 n$_{0}$) & ($\sim$ 2.0 n$_{0}$) & ($\sim$ 2.5 n$_{0}$) & ($\sim$ 3.0 n$_{0}$)\\ \hline
GDFMX   & 0.884 & 0.971 & 0.921 & 0.827 \\
DDB     & 0.941 & 0.939 & 0.847 & 0.694 \\
DDBy    & 0.955 & 0.941 & 0.835 & 0.683 \\
DDH     & 0.869 & 0.963 & 0.802 & 0.630 \\
DDHy    & 0.922 & 0.955 & 0.772 & 0.564 \\
RMF-NL  & 0.879 & 0.937 & 0.845 & 0.694 \\
\hline
\end{tabular}
 \label{tab:cor2rho0}
\end{table}

\clearpage

\begin{figure}
    \centering
    \includegraphics[width=0.4\linewidth]{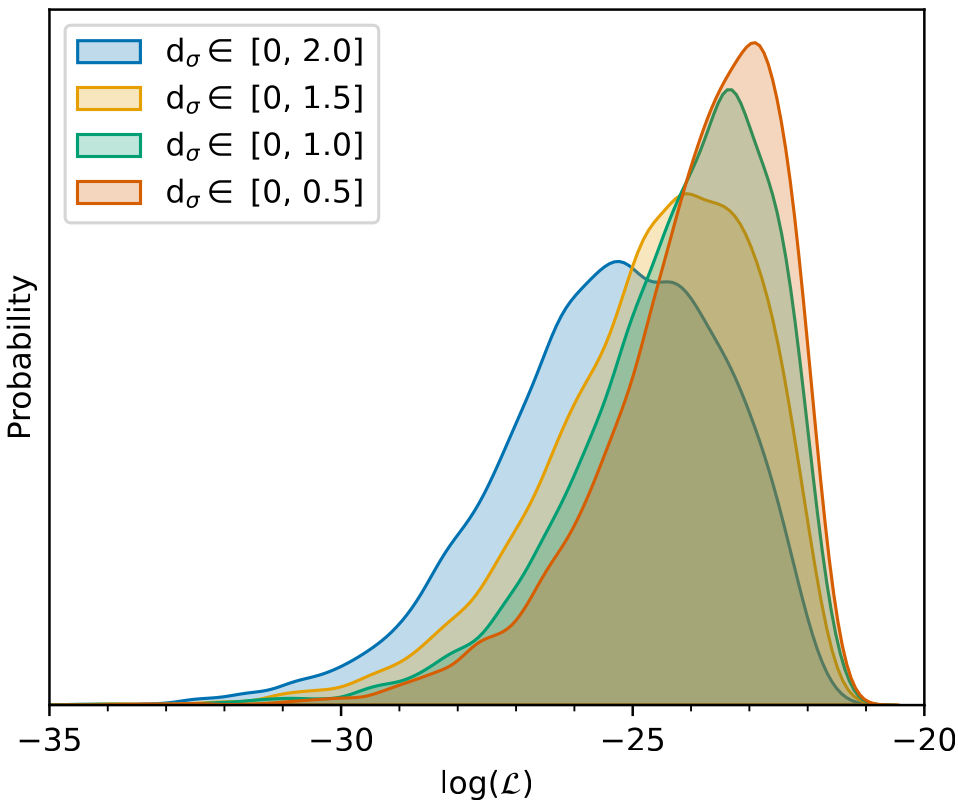}
    \caption{The total likelihood distribution of the DDH model for different prior ranges of $d_\sigma$.}
    \label{fig:ddh_prior_fix}
\end{figure}

The following tables give the priors and the posteriors of the model parameters considered in the present study.
% Table Fix Prior %

\begin{table}[ht!]
\centering
\caption{Prior ranges for all models: RMF-NL, DDH, DDHy, DDB, DDBy, GDFMX.}
\label{tab:prior}
\begin{tabular}{cc cc cc}
\hline
\multicolumn{2}{c}{RMF-NL} & \multicolumn{2}{c}{DDH} & \multicolumn{2}{c}{DDHy} \\
Quantity & Range & Quantity & Range & Quantity & Range \\
\hline
\StretchedColumn{5cm}{1cm}{
\vspace{0.35\baselineskip}
$\Gamma_{\sigma}$\\[\stretch{1}]
$\Gamma_{\omega}$\\[\stretch{1}]
$\Gamma_{\rho}$\\[\stretch{1}]
$\kappa$\\[\stretch{1}]
$\lambda_0$\\[\stretch{1}]
$\zeta$\\[\stretch{1}]
$\Lambda_w$\\[\stretch{1}]
\vspace{-0.5\baselineskip}
} 
& \StretchedColumn{5cm}{2.5cm}{
\vspace{0.35\baselineskip}
$\BracketRange{6.5}{15.5}$\\[\stretch{1}]
$\BracketRange{6.5}{15.5}$\\[\stretch{1}]
$\BracketRange{5.5}{16.5}$\\[\stretch{1}]
$\BracketRange{0.0048}{0.086}$\\[\stretch{1}]
$\BracketRange{-0.05}{0.05}$\\[\stretch{1}]
$\BracketRange{0.00}{0.04}$\\[\stretch{1}]
$\BracketRange{0.00}{0.15}$\\[\stretch{1}]
\vspace{-0.5\baselineskip}
} 
& \StretchedColumn{5cm}{1cm}{
\vspace{0.35\baselineskip}
$\Gamma_{\sigma,0}$\\[\stretch{1}]
$\Gamma_{\omega,0}$\\[\stretch{1}]
$\Gamma_{\rho,0}$\\[\stretch{1}]
$d_{\sigma}$\\[\stretch{1}]
$a_{\sigma}$\\[\stretch{1}]
$d_{\omega}$\\[\stretch{1}]
$a_{\rho}$\\[\stretch{1}]
\vspace{-0.5\baselineskip}
} 
& \StretchedColumn{5cm}{2.5cm}{
\vspace{0.35\baselineskip}
$\BracketRange{6.5}{14.0}$\\[\stretch{1}]
$\BracketRange{6.5}{15.0}$\\[\stretch{1}]
$\BracketRange{5.0}{17.0}$\\[\stretch{1}]
$\BracketRange{0.0}{1.0}$\\[\stretch{1}]
$\BracketRange{1.0}{2.0}$\\[\stretch{1}]
$\BracketRange{0.0}{2.0}$\\[\stretch{1}]
$\BracketRange{0.0}{1.5}$\\[\stretch{1}]
\vspace{-0.5\baselineskip}
} 
& \StretchedColumn{5cm}{1cm}{
\vspace{0.16\baselineskip}
$\Gamma_{\sigma,0}$\\[\stretch{1}]
$\Gamma_{\omega,0}$\\[\stretch{1}]
$\Gamma_{\rho,0}$\\[\stretch{1}]
$d_{\sigma}$\\[\stretch{1}]
$a_{\sigma}$\\[\stretch{1}]
$d_{\omega}$\\[\stretch{1}]
$a_{\rho}$\\[\stretch{1}]
$y$\\[\stretch{1}]
\vspace{-0.5\baselineskip}
} 
& \StretchedColumn{5cm}{2.5cm}{
\vspace{0.16\baselineskip}
$\BracketRange{6.5}{14.0}$\\[\stretch{1}]
$\BracketRange{6.5}{15.0}$\\[\stretch{1}]
$\BracketRange{5.0}{17.0}$\\[\stretch{1}]
$\BracketRange{0.0}{1.0}$\\[\stretch{1}]
$\BracketRange{1.0}{2.0}$\\[\stretch{1}]
$\BracketRange{0.0}{2.0}$\\[\stretch{1}]
$\BracketRange{0.0}{1.5}$\\[\stretch{1}]
$\BracketRange{0.0}{1.0}$\\[\stretch{1}]
\vspace{-0.5\baselineskip}
} \\
\hline
\multicolumn{2}{c}{DDB} & \multicolumn{2}{c}{DDBy} & \multicolumn{2}{c}{GDFMX} \\
Quantity & Range & Quantity & Range & Quantity & Range \\
\hline
\StretchedColumn{6.5cm}{1cm}{
\vspace{1.0\baselineskip}
$a_{\sigma}$\\[\stretch{1}]
$a_{\omega}$\\[\stretch{1}]
$a_{\rho}$\\[\stretch{1}]
$\Gamma_{\sigma}$\\[\stretch{1}]
$\Gamma_{\omega}$\\[\stretch{1}]
$\Gamma_{\rho}$\\[\stretch{1}]
\vspace{-0.5\baselineskip}
} 
& \StretchedColumn{6.5cm}{2.5cm}{
\vspace{1.0\baselineskip}
$\BracketRange{0.0}{0.3}$\\[\stretch{1}]
$\BracketRange{0.0}{0.3}$\\[\stretch{1}]
$\BracketRange{0.0}{1.5}$\\[\stretch{1}]
$\BracketRange{6.5}{13.5}$\\[\stretch{1}]
$\BracketRange{7.5}{14.5}$\\[\stretch{1}]
$\BracketRange{5.0}{16.0}$\\[\stretch{1}]
\vspace{-0.5\baselineskip}
} 
& \StretchedColumn{6.5cm}{1cm}{
\vspace{0.72\baselineskip}
$a_{\sigma}$\\[\stretch{1}]
$a_{\omega}$\\[\stretch{1}]
$a_{\rho}$\\[\stretch{1}]
$\Gamma_{\sigma}$\\[\stretch{1}]
$\Gamma_{\omega}$\\[\stretch{1}]
$\Gamma_{\rho}$\\[\stretch{1}]
$y$\\[\stretch{1}]
\vspace{-0.5\baselineskip}
} 
& \StretchedColumn{6.5cm}{2.5cm}{
\vspace{0.72\baselineskip}
$\BracketRange{0.0}{0.3}$\\[\stretch{1}]
$\BracketRange{0.0}{0.3}$\\[\stretch{1}]
$\BracketRange{0.0}{1.5}$\\[\stretch{1}]
$\BracketRange{6.5}{13.5}$\\[\stretch{1}]
$\BracketRange{7.5}{14.5}$\\[\stretch{1}]
$\BracketRange{5.0}{16.0}$\\[\stretch{1}]
$\BracketRange{0.0}{1.0}$\\[\stretch{1}]
\vspace{-0.5\baselineskip}
} 
& \StretchedColumn{6.5cm}{1cm}{
\vspace{-0.01\baselineskip}
$a_{\sigma}$\\[\stretch{1}]
$a_{\omega}$\\[\stretch{1}]
$a_{\rho}$\\[\stretch{1}]
$b_{\sigma}$\\[\stretch{1}]
$b_{\omega}$\\[\stretch{1}]
$b_{\rho}$\\[\stretch{1}]
$c_{\sigma}$\\[\stretch{1}]
$c_{\omega}$\\[\stretch{1}]
$c_{\rho}$\\[\stretch{1}]
$d_{\sigma}$\\[\stretch{1}]
$d_{\omega}$\\[\stretch{1}]
$d_{\rho}$\\[\stretch{1}]
\vspace{-0.5\baselineskip}
} 
& \StretchedColumn{6.5cm}{2.5cm}{
\vspace{-0.01\baselineskip}
$\BracketRange{5.5}{10.5}$\\[\stretch{1}]
$\BracketRange{8.0}{14.0}$\\[\stretch{1}]
$\BracketRange{0.0}{12.0}$\\[\stretch{1}]
$\BracketRange{1.0}{6.0}$\\[\stretch{1}]
$\BracketRange{1.0}{7.0}$\\[\stretch{1}]
$\BracketRange{1.0}{20.0}$\\[\stretch{1}]
$\BracketRange{0.0}{2.0}$\\[\stretch{1}]
$\BracketRange{0.0}{2.0}$\\[\stretch{1}]
$\BracketRange{0.0}{2.0}$\\[\stretch{1}]
$\BracketRange{0.0}{40.0}$\\[\stretch{1}]
$\BracketRange{0.0}{9.0}$\\[\stretch{1}]
$\BracketRange{0.0}{6.0}$\\[\stretch{1}]
\vspace{-0.5\baselineskip}
} \\
\hline
\end{tabular}
\end{table}

\begin{table}[ht!]
\centering
\caption{The median and 90\% CI for the posterior distributions of all models: RMF-NL, DDH, DDHy, DDB, DDBy, GDFMX.}
\label{tab:posterior}
\begin{tabular}{cc cc cc}
\hline
\multicolumn{2}{c}{RMF-NL} & \multicolumn{2}{c}{DDH} & \multicolumn{2}{c}{DDHy} \\
Quantity & Value & Quantity & Value & Quantity & Value \\
\hline
\StretchedColumn{5cm}{1cm}{
\vspace{0.2\baselineskip}
$\Gamma_{\sigma}^{\phantom{0}}$\\[\stretch{1}]
$\Gamma_{\omega}^{\phantom{0}}$\\[\stretch{1}]
$\Gamma_{\rho}^{\phantom{0}}$\\[\stretch{1}]
$\kappa_{\phantom{0}}^{\phantom{0}}$\\[\stretch{1}]
$\lambda_0^{\phantom{0}}$\\[\stretch{1}]
$\zeta_{\phantom{0}}^{\phantom{0}}$\\[\stretch{1}]
$\Lambda_w^{\phantom{0}}$\\[\stretch{1}]
\vspace{-0.5\baselineskip}
}
& \StretchedColumn{5cm}{2.5cm}{
\vspace{0.2\baselineskip}
$\phantom{0}8.231^{+0.604}_{-0.413}$\\[\stretch{1}]
$\phantom{0}9.453^{+1.113}_{-0.721}$\\[\stretch{1}]
$11.132^{+0.818}_{-0.871}$\\[\stretch{1}]
$\phantom{0}0.053^{+0.022}_{-0.022}$\\[\stretch{1}]
$\hspace{-0.4cm}\phantom{0}-0.008^{+0.048}_{-0.030}$\\[\stretch{1}]
$\phantom{0}0.004^{+0.009}_{-0.004}$\\[\stretch{1}]
$\phantom{0}0.085^{+0.048}_{-0.036}$\\[\stretch{1}]
\vspace{-0.5\baselineskip}
}
& \StretchedColumn{5cm}{1cm}{
\vspace{0.2\baselineskip}
$\Gamma_{\sigma}^{\phantom{0}}$\\[\stretch{1}]
$\Gamma_{\omega}^{\phantom{0}}$\\[\stretch{1}]
$\Gamma_{\rho}^{\phantom{0}}$\\[\stretch{1}]
$a_{\sigma}^{\phantom{0}}$\\[\stretch{1}]
$d_{\sigma}^{\phantom{0}}$\\[\stretch{1}]
$d_{\omega}^{\phantom{0}}$\\[\stretch{1}]
$d_{\rho}^{\phantom{0}}$\\[\stretch{1}]
\vspace{-0.5\baselineskip}
}
& \StretchedColumn{5cm}{2.5cm}{
\vspace{0.2\baselineskip}
$\phantom{0}8.944^{+1.974}_{-0.720}$\\[\stretch{1}]
$10.713^{+2.933}_{-1.145}$\\[\stretch{1}]
$\phantom{0}7.578^{+0.781}_{-1.181}$\\[\stretch{1}]
$\phantom{0}1.323^{+0.100}_{-0.088}$\\[\stretch{1}]
$\phantom{0}0.764^{+0.200}_{-0.289}$\\[\stretch{1}]
$\phantom{0}0.605^{+0.387}_{-0.359}$\\[\stretch{1}]
$\phantom{0}0.409^{+0.257}_{-0.205}$\\[\stretch{1}]
\vspace{-0.5\baselineskip}
}
& \StretchedColumn{5cm}{1cm}{
\vspace{0.01\baselineskip}
$\Gamma_{\sigma}^{\phantom{0}}$\\[\stretch{1}]
$\Gamma_{\omega}^{\phantom{0}}$\\[\stretch{1}]
$\Gamma_{\rho}^{\phantom{0}}$\\[\stretch{1}]
$a_{\sigma}^{\phantom{0}}$\\[\stretch{1}]
$d_{\sigma}^{\phantom{0}}$\\[\stretch{1}]
$d_{\omega}^{\phantom{0}}$\\[\stretch{1}]
$d_{\rho}^{\phantom{0}}$\\[\stretch{1}]
$y_{\phantom{0}}^{\phantom{0}}$\\[\stretch{1}]
\vspace{-0.5\baselineskip}
}
& \StretchedColumn{5cm}{2.5cm}{
\vspace{0.01\baselineskip}
$\phantom{0}8.727^{+1.626}_{-0.671}$\\[\stretch{1}]
$10.374^{+2.460}_{-1.072}$\\[\stretch{1}]
$\phantom{0}7.714^{+0.718}_{-0.963}$\\[\stretch{1}]
$\phantom{0}1.321^{+0.091}_{-0.088}$\\[\stretch{1}]
$\phantom{0}0.754^{+0.208}_{-0.256}$\\[\stretch{1}]
$\phantom{0}0.551^{+0.398}_{-0.316}$\\[\stretch{1}]
$\phantom{0}0.685^{+0.610}_{-0.387}$\\[\stretch{1}]
$\phantom{0}0.511^{+0.415}_{-0.335}$\\[\stretch{1}]
\vspace{-0.5\baselineskip}
}
\\\hline

\multicolumn{2}{c}{DDB} & \multicolumn{2}{c}{DDBy} & \multicolumn{2}{c}{GDFMX} \\
Quantity & Value & Quantity & Value & Quantity & Value \\
\hline
\StretchedColumn{7.0cm}{1cm}{
\vspace{1.0\baselineskip}
$\Gamma_{\sigma}^{\phantom{0}}$\\[\stretch{1}]
$\Gamma_{\omega}^{\phantom{0}}$\\[\stretch{1}]
$\Gamma_{\rho}^{\phantom{0}}$\\[\stretch{1}]
$a_{\sigma}^{\phantom{0}}$\\[\stretch{1}]
$a_{\omega}^{\phantom{0}}$\\[\stretch{1}]
$a_{\rho}^{\phantom{0}}$\\[\stretch{1}]
\vspace{-0.5\baselineskip}
}
& \StretchedColumn{7.0cm}{2.5cm}{
\vspace{1.0\baselineskip}
$\phantom{0}9.014^{+1.086}_{-0.831}$\\[\stretch{1}]
$10.755^{+1.631}_{-1.317}$\\[\stretch{1}]
$\phantom{0}7.498^{+0.718}_{-0.926}$\\[\stretch{1}]
$\phantom{0}0.092^{+0.031}_{-0.023}$\\[\stretch{1}]
$\phantom{0}0.061^{+0.060}_{-0.052}$\\[\stretch{1}]
$\phantom{0}0.574^{+0.166}_{-0.137}$\\[\stretch{1}]
\vspace{-0.5\baselineskip}
}
& \StretchedColumn{7.0cm}{1cm}{
\vspace{0.7\baselineskip}
$\Gamma_{\sigma}^{\phantom{0}}$\\[\stretch{1}]
$\Gamma_{\omega}^{\phantom{0}}$\\[\stretch{1}]
$\Gamma_{\rho}^{\phantom{0}}$\\[\stretch{1}]
$a_{\sigma}^{\phantom{0}}$\\[\stretch{1}]
$a_{\omega}^{\phantom{0}}$\\[\stretch{1}]
$a_{\rho}^{\phantom{0}}$\\[\stretch{1}]
$y_{\phantom{0}}^{\phantom{0}}$\\[\stretch{1}]
\vspace{-0.5\baselineskip}
}
& \StretchedColumn{7.0cm}{2.5cm}{
\vspace{0.7\baselineskip}
$\phantom{0}8.849^{+1.148}_{-0.826}$\\[\stretch{1}]
$10.501^{+1.736}_{-1.320}$\\[\stretch{1}]
$\phantom{0}7.626^{+0.728}_{-0.879}$\\[\stretch{1}]
$\phantom{0}0.092^{+0.031}_{-0.022}$\\[\stretch{1}]
$\phantom{0}0.056^{+0.063}_{-0.049}$\\[\stretch{1}]
$\phantom{0}0.867^{+0.523}_{-0.324}$\\[\stretch{1}]
$\phantom{0}0.586^{+0.362}_{-0.307}$\\[\stretch{1}]
\vspace{-0.5\baselineskip}
}
& \StretchedColumn{7.0cm}{1cm}{
\vspace{-0.0\baselineskip}
$a_{\sigma}^{\phantom{0}}$\\[\stretch{1}]
$a_{\omega}^{\phantom{0}}$\\[\stretch{1}]
$a_{\rho}^{\phantom{0}}$\\[\stretch{1}]
$b_{\sigma}^{\phantom{0}}$\\[\stretch{1}]
$b_{\omega}^{\phantom{0}}$\\[\stretch{1}]
$b_{\rho}^{\phantom{0}}$\\[\stretch{1}]
$c_{\sigma}^{\phantom{0}}$\\[\stretch{1}]
$c_{\omega}^{\phantom{0}}$\\[\stretch{1}]
$c_{\rho}^{\phantom{0}}$\\[\stretch{1}]
$d_{\sigma}^{\phantom{0}}$\\[\stretch{1}]
$d_{\omega}^{\phantom{0}}$\\[\stretch{1}]
$d_{\rho}^{\phantom{0}}$\\[\stretch{1}]
\vspace{-0.5\baselineskip}
}
& \StretchedColumn{7.0cm}{2.5cm}{
\vspace{-0.0\baselineskip}
$\phantom{0}7.584^{+0.684}_{-0.858}$\\[\stretch{1}]
$\phantom{0}9.240^{+0.424}_{-0.553}$\\[\stretch{1}]
$\phantom{0}2.366^{+1.025}_{-1.665}$\\[\stretch{1}]
$\phantom{0}2.643^{+1.305}_{-0.726}$\\[\stretch{1}]
$\phantom{0}2.448^{+2.121}_{-1.158}$\\[\stretch{1}]
$\phantom{0}3.959^{+1.811}_{-1.767}$\\[\stretch{1}]
$\phantom{0}1.038^{+0.710}_{-0.410}$\\[\stretch{1}]
$\phantom{0}1.395^{+0.469}_{-0.670}$\\[\stretch{1}]
$\phantom{0}1.100^{+0.736}_{-0.635}$\\[\stretch{1}]
$\phantom{0}0.243^{+0.733}_{-0.196}$\\[\stretch{1}]
$\phantom{0}0.922^{+2.774}_{-0.842}$\\[\stretch{1}]
$\phantom{0}0.483^{+3.085}_{-0.457}$\\[\stretch{1}]
\vspace{-0.5\baselineskip}
}
\\\hline
\end{tabular}
\end{table}

% Table Fix Posterior %
\begin{figure*}
    \centering
    \includegraphics[width=0.23\linewidth]{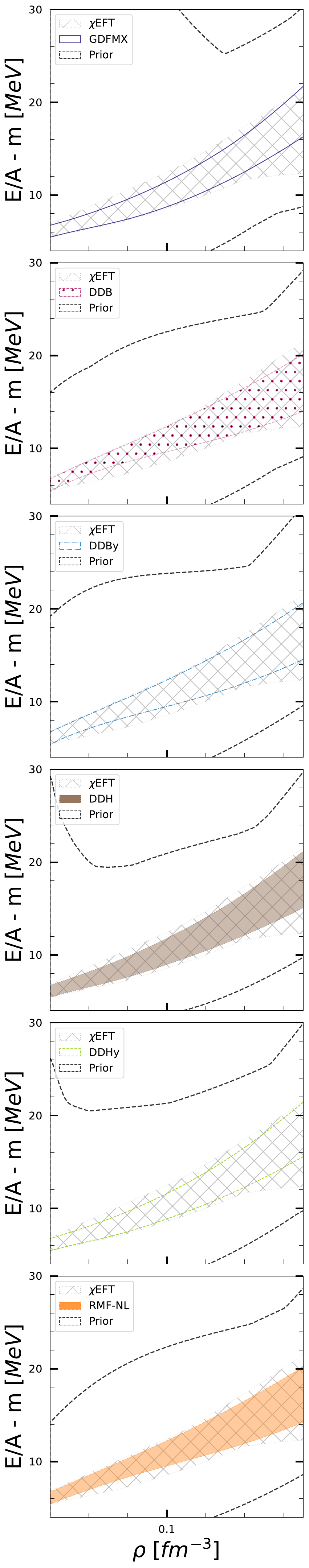}
    \includegraphics[width=0.23346\linewidth]{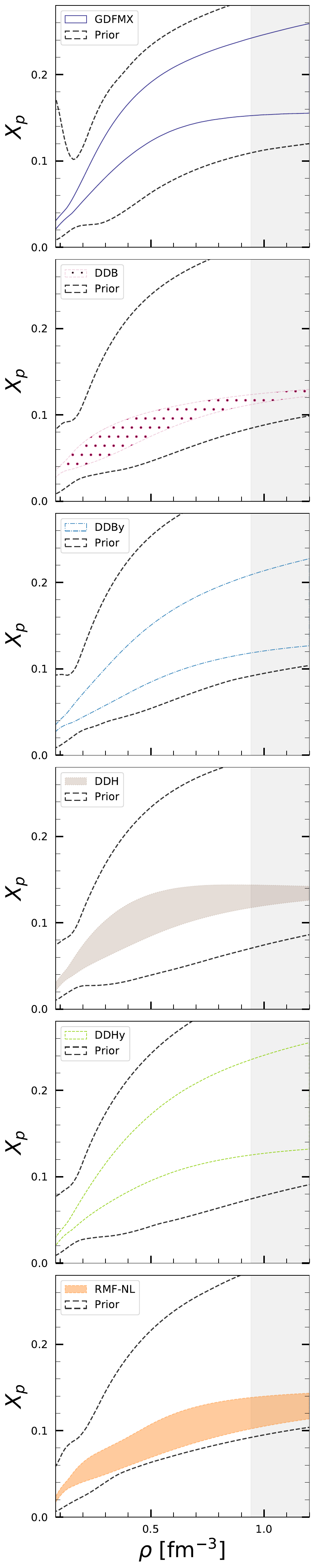}
    \includegraphics[width=0.2373\linewidth]{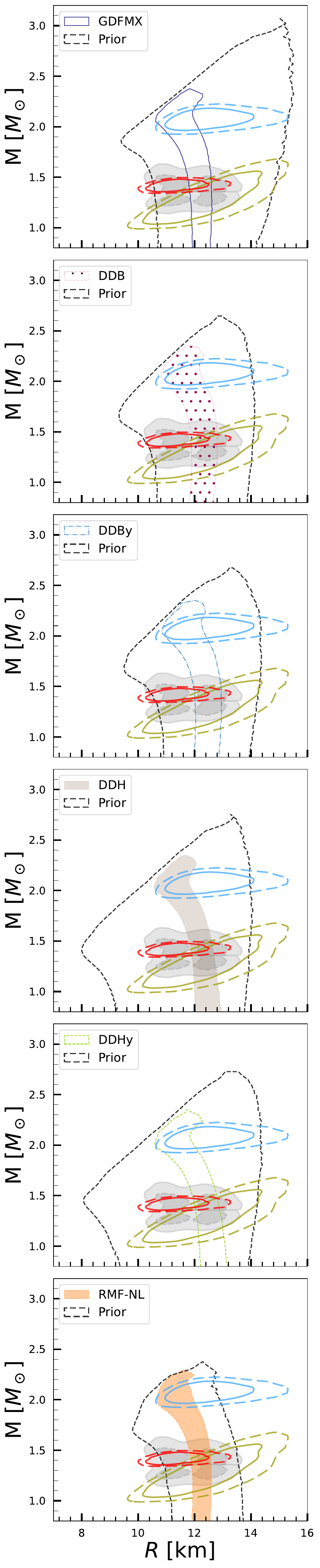}
    \caption{Full prior domain (dashed black lines) and 90\% credible intervals (shaded regions) of the posterior distributions for six different models. Left column: pure neutron matter energy per particle ($E/A - m$) as a function of baryon density $\rho$. Middle column: proton fraction ($X_p$) as a function of density. Right column: mass–radius relations for neutron stars.}
    \label{fig:prior_compare}
\end{figure*}

\begin{figure*}
    \centering
    \includegraphics[width=0.75\linewidth]{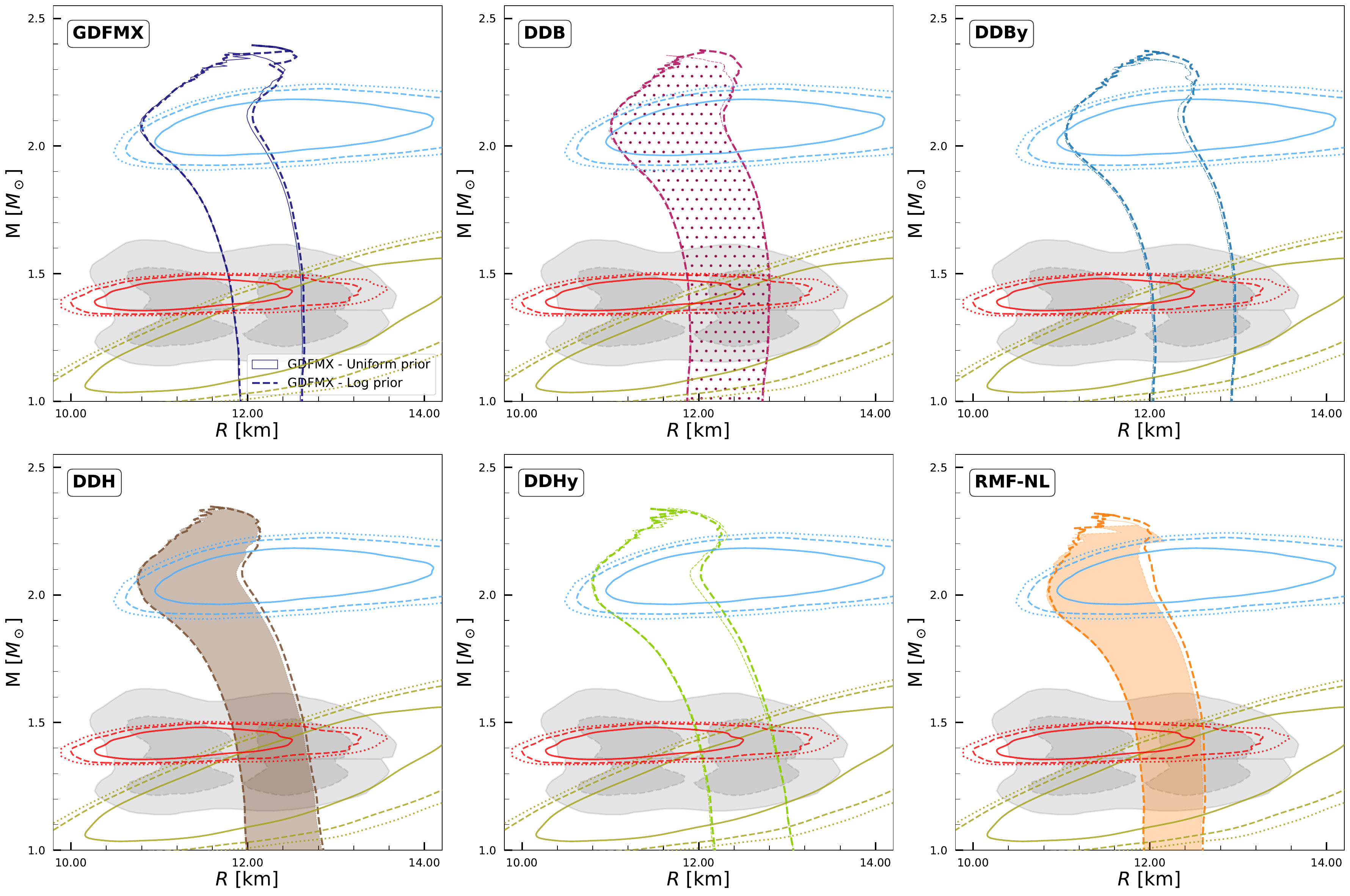} 
    \caption{Mass-radius relations for six EoS models comparing uniform prior (filled regions) and logarithmic prior (dashed contours) on the pQCD renormalization scale X. The pQCD constraint is applied at $\rho$ = 1.2~fm$^{-3}$. Observational constraints from PSR J0030+0451 (yellow), PSR J0740+6620 (blue), PSR J0437-4715 (red), and GW170817 (gray shaded regions for 50\% and 90\% credible intervals) are shown for reference. Shaded regions show 90\% credible intervals for each model. The logarithmic prior gives a larger weight to the less constraining condition (X=1). As a consequence, the uniform prior is more restrictive.}
    \label{fig:pQCD prior1}
\end{figure*}

\clearpage
\bibliography{library}{}
\bibliographystyle{aasjournal}
\end{document}